\documentclass[12pt,letterpaper]{article}

\usepackage{textcomp}
\usepackage{chetnew}
\usepackage{tikz}
\usepackage{amssymb}
\usepackage{comment}
\usepackage[utf8]{inputenc}

\usepackage{xcolor}
\usepackage{tikz}

\newcommand{\bx}{{\bf x}}

\renewcommand{\tilde}{\widetilde}

\newcommand{\CB}{\mathcal{B}}

\newcommand{\CQ}{\mathcal{Q}}
\newcommand{\CH}{\mathcal{H}}

\newcommand{\CC}{\mathcal{C}}
\newcommand{\CO}{\mathcal{O}}
\newcommand{\CT}{\mathcal{T}}

\newcommand{\CI}{\mathcal{I}}
\newcommand{\CN}{\mathcal{N}}
\newcommand{\CS}{\mathcal{S}}
\newcommand{\CM}{\mathcal{M}}

\preprint{QMUL-PH-17-XX}

\title{\vspace*{-1.5cm} On Irregular Singularity Wave Functions\\[2mm] and Superconformal Indices}

\author{Matthew Buican$^{\diamondsuit, 1}$ and Takahiro Nishinaka$^{\clubsuit, 2,3}$}

\affiliation{\smallskip$^1$ CRST and School of Physics and Astronomy\\
Queen Mary University of London, London E1 4NS, UK\\ \smallskip$^2$Yukawa Institute for Theoretical Physics\\ Kyoto University, Kyoto 606-8502, Japan\\ \smallskip$^3$ Department of Physical Sciences, College of Science and Engineering \\ Ritsumeikan University, Shiga 525-8577 Japan
\emails{$^{\diamondsuit}$m.buican@qmul.ac.uk, $^{\clubsuit}$nishinak@fc.ritsumei.ac.jp}}

\abstract{We generalize, in a manifestly Weyl-invariant way, our previous expressions for irregular singularity wave functions in two-dimensional $SU(2)$ $q$-deformed Yang-Mills theory to $SU(N)$. As an application, we give closed-form expressions for the Schur indices of all $(A_{N-1}, A_{N(n-1)-1})$ Argyres-Douglas (AD) superconformal field theories (SCFTs), thus completing the computation of these quantities for the $(A_N, A_M)$ SCFTs. With minimal effort, our wave functions also give new Schur indices of various infinite sets of \lq\lq Type IV" AD theories. We explore the discrete symmetries of these indices and also show how highly intricate renormalization group (RG) flows from isolated theories and conformal manifolds in the ultraviolet to isolated theories and (products of) conformal manifolds in the infrared are encoded in these indices. We compare our flows with dimensionally reduced flows via a simple \lq\lq monopole vev RG" formalism. Finally, since our expressions are given in terms of concise Lie algebra data, we speculate on extensions of our results that might be useful for probing the existence of hypothetical SCFTs based on other Lie algebras. We conclude with a discussion of some open problems.}

\date{May 2017}

\begin{document}
\setcounter{tocdepth}{2}

\maketitle

\toc

\section{Introduction}
Inspired by constructions of certain four-dimensional (4D) superconformal indices as correlators in 2D topological field theory (TFT) on a (punctured) Riemann surface $\CC$ \cite{Gadde:2011ik}, we proposed a generalization in \cite{Buican:2015ina} that leads to closed-form expressions for the Schur limit of the superconformal indices of two infinite sets of Argyres-Douglas (AD) theories that arise from twisted compactifications of the 6D $A_1$ $(2,0)$ theory on $\CC$---the so-called $(A_1, A_{2n-3})$ and $(A_1, D_{2n})$ superconformal field theories (SCFTs).\footnote{See \cite{Cecotti:2010fi,Xie:2012hs} for nomenclature and the classic references \cite{Argyres:1995jj,Argyres:1995xn,Eguchi:1996vu} for the original constructions of these theories as endpoints of renormalization group flows from $\CN=2$ gauge theories.} In addition to giving exact information about non-trivial sectors of these theories (the so-called Schur operators \cite{Gadde:2011uv,Beem:2013sza}) and characterizing new states in 2D $SU(2)$ $q$-deformed Yang-Mills (see \cite{Cordes:1994fc} for a review and, e.g.,  \cite{deHaro:2006uvl,Kimura:2008gs,Szabo:2013vva} for other recent developments), these indices contain a surprise: they encode information about the $\CN=2$ chiral operators\footnote{By $\CN=2$ chiral operators, we mean operators annihilated by all the anti-chiral $\CN=2$ Poincar\'e supercharges.} parameterizing the Coulomb branch even though $\CN=2$ chiral operators do not contribute directly to the Schur index \cite{Buican:2015hsa}. These results may point to the existence of a deeper structure at work in 4D $\CN=2$ SCFTs (see \cite{Fredrickson:2017yka} for interesting recent progress in this direction). In fact, Coulomb branch physics is at the heart of a complementary approach to computing these indices via BPS state counting \cite{Cordova:2015nma} (building on results in \cite{Iqbal:2012xm}).

More recently, many papers have appeared that include generalizations to other classes of Argyres-Douglas theories\footnote{We define an Argyres-Douglas theory to be any $\CN=2$ SCFT with at least one $\CN=2$ chiral operator of non-integer scaling dimension.} and other limits of the superconformal index \cite{Buican:2015tda,Song:2015wta,Cecotti:2015lab,Xie:2016evu} as well as to the full superconformal index \cite{Maruyoshi:2016tqk,Maruyoshi:2016aim,Agarwal:2016pjo} (and also to minimal interacting deformations of AD theories \cite{Xie:2016hny,Buican:2016hnq}). However, many interesting Argyres-Douglas theories remain to be explored, and various aspects of the structure underlying these theories remain to be uncovered (see \cite{Cordova:2016uwk,Cordova:2017ohl,Cordova:2017mhb} for interesting recent progress).

In this paper, we generalize our discussion in \cite{Buican:2015ina} and propose the following simple wave functions for certain irregular punctures in $SU(N)$ $q$-deformed Yang-Mills theory (thus generalizing our earlier results from $N=2$ to all $N\ge2$)
\begin{align}
\tilde{f}_{R}^{(n)}(q;{\bf x}) &= \prod_{k=1}^\infty\left(\frac{1}{1-q^k}\right)^{N-1}q^{nC_2(R)}\text{Tr}_R\left[q^{-\frac{n}{2}F^{ij}h_ih_j}\prod_{i=1}^{N-1}(x_1\cdots x_i)^{h_i}\right]~,
\label{eq:wf1}
\end{align}
where $n\ge2$ is an integer, $q$ is a fugacity, $R$ is an irreducible representation of $A_{N-1}$ with quadratic Casimir $C(R)$ and Cartans $h_i$ (in the Chevalley basis\footnote{Namely, the ``rasing'' and ``lowering'' generators, $e_i$ and $f_i$, satisfy $[h_i,\, e_j] = A_{ji} e_j,\, [h_i, f_j] = -A_{ji}f_j$ and $[e_i,f_j] = \delta_{ij}h_j$ with the Cartan matrix, $A_{ij}$.}), ${\bf x}=(x_1,\cdots,x_{N-1})$ are flavor fugacities, and the factor $F^{ij}$ is the quadratic form matrix, i.e., the inverse of the Cartan matrix
\begin{align}\label{Cartinv}
F
= \frac{1}{N}\left[
\begin{array}{ccccc}
N-1 & N-2 & N-3 & \cdots & 1\\
N-2 & 2(N-2)& 2(N-3)& \cdots & 2\\
N-3 & 2(N-3) & 3(N-3) & \cdots & 3\\
\vdots &\vdots& \vdots& \ddots & \vdots \\
1 & 2& 3&\cdots& N-1\\
\end{array}
\right]~.
\end{align}
In particular, this wave function can be used to construct Schur indices for the $(A_{N-1}, A_{N(n-1)-1})$ Argyres-Douglas theories
\begin{align}
\mathcal{I}_{(A_{N-1},A_{N(n-1)-1})}(q;{\bf x}) = \sum_{R}C_R(q)\tilde{f}^{(n)}_R(q;{\bf x})~,
\label{eq:general} 
\end{align}
where the sum is taken over all irreducible representations of $A_{N-1}$, and the coefficients, $C_R$, take the form
\begin{align}\label{topological}
C_{R}(q) = \frac{\prod_{k=1}^{N-1}(1-q^k)^{N-k}}{(q;q)_\infty^{N-1}}\text{dim}_q R = \frac{\prod_{k=1}^{N-1}(1-q^k)^{N-k}}{(q;q)_\infty^{N-1}}\chi_R^{su(N)}(q^{-\frac{N-1}{2}},q^{-\frac{N-3}{2}},\cdots,q^{\frac{N-1}{2}}) ~,
\end{align}
as conjectured in \cite{Gadde:2011ik}.\footnote{These coefficients are fixed by the rank of the $q$-deformed Yang-Mills theory and the topology of the surface on which it lives. As we will review below, both these properties are in turn fixed---up to duality---by the particular AD theory we consider.} Here, our convention for the character is such that
$\chi^{su(N)}_R(x_1,\cdots,x_{N}) \equiv \text{Tr}_{R} \left[\prod_{i=1}^{N-1}(x_1\cdots x_i)^{h_i}\right]$ with $x_{N} \equiv (x_1\cdots x_{N-1})^{-1}$.\footnote{\label{foot:ell}An equivalent expression is
\begin{align}
\chi^{su(N)}_R(x_1,\cdots,x_{N}) = \frac{\det (x_j^{\ell_i+N-i})}{\det (x_j^{N-i})}~,
\end{align}
where $\ell_1,\,\cdots,\ell_{N}$ are given by $\ell_i = \sum_{j=i}^{N-1}\lambda_j$, in terms of the Dynkin labels, $(\lambda_1,\cdots\lambda_{N-1})$, of the representaiton $R$.}

Therefore, in addition to providing a description of new states in $SU(N)$ $q$-deformed Yang-Mills theory, our expression in \eqref{eq:wf1} can be used to construct closed-form expressions for Schur indices of a doubly infinite set of strongly interacting SCFTs. This proposal completes the construction of all such indices for Argyres-Douglas theories of type $(A_N, A_M)$. Indeed, these indices have not been previously constructed for theories of type $(A_{N-1}, A_{N(n-1)-1})$ with $N>2$ (with the exception of the $(A_3, A_3)$ and $(A_2, A_5)$ cases for which expressions involving integrals over gauge groups exist \cite{Buican:2015ina,Buican:2015tda}, but no simple sum of the type in \eqref{eq:wf1} has been found\footnote{One may also try to infer additional integral expressions for some of these theories using the methods described in \cite{Xie:2016uqq,Xie:2017vaf} and results in the existing literature.}).

One interesting aspect of the $(A_{N-1}, A_{N(n-1)-1})$ theories with $N>2$ is that they typically have exactly marginal deformations (if $n=2$, there are $N-3$ such deformations, and, if $n>2$, there are $N-2$ exactly marginal deformations). While the index is an invariant of the resulting conformal manifolds, the $S$-duality groups (see the interesting recent discussion in  \cite{Caorsi:2016ebt}) act on the index through discrete symmetries. Our compact expressions for the Schur indices make it possible to explore the discrete symmetries of the index efficiently.

Moreover, as we will see in detail, our formulae encode a highly non-trivial set of renormalization group (RG) flows that typically take conformal manifolds in the ultraviolet (UV) and often map them to products of conformal manifolds in the IR along with various isolated factors. While we leave a deeper exploration of such RG flows and the laws they obey to future work, we develop a simple \lq\lq monopole vev RG flow" formalism to study these flows in the theories related by mirror symmetry to the $S^1$ reductions of our AD theories of interest (we explain why the reduction along the circle commutes with the RG flow).

Another aspect of our proposal is that it immediately gives us an infinite set of new superconformal indices for free. Indeed, simply by including already-existing expressions for wave functions corresponding to an additional regular puncture in the $SU(N)$ $q$-deformed Yang-Mills theory, we generate Schur indices for infinitely many co-called \lq\lq type IV" Argyres-Douglas theories \cite{Xie:2012hs,Xie:2013jc}
\begin{align}\label{2puncture}
\mathcal{I}_{(I_{N,N(n-1)},R_Y)}(q;{\bf x};{\bf y}) = \sum_{R}\tilde{f}^{(n)}_R(q;{\bf x})f_R^{Y}(q;{\bf y})~, 
\end{align}
where
\begin{align}\label{Regular}
f_{R}^{Y}(q;{\bf y}) &= P.E.\left[\frac{q}{1-q}\chi^{\rho(Y)}(q,{\bf y})\right]\chi^{\text{SU}(N)}_R\left(\Lambda^Y({\bf y})\right)~,
\end{align}
 ${\bf y}$ is the flavor fugacity associated with the regular puncture, $\Lambda^Y({\bf y})$ is the set of $N-1$ fugacities corresponding to the Young diagram $Y$ (see \cite{Gadde:2011ik}), $\chi^{\rho(Y)}(q,{\bf y})$ is a polynomial in $q$ determined by $Y$, and 
\begin{equation}
P.E.\left[G(a_1,\cdots,a_p)\right]\equiv\exp\left[\sum_{n=1}^{\infty}{1\over n}G(a_1^n,\cdots, a_p^n)\right]~,
\end{equation}
for any function of the fugacities, $G$.

Finally, the fact that our expressions are written in terms of simple Lie algebra data makes it tempting to speculate about possible generalizations of our expressions beyond $A_{N-1}$ (and perhaps beyond the theories one can engineer from compactifications of the $(2,0)$ theory on a Riemann surface, $\CC$). In section \ref{speculation}, we explore these ideas further.

The plan of this paper is as follows. In the next section, we briefly review basic aspects of the $(A_{N-1}, A_{N(n-1)-1})$ theories including their Seiberg-Witten curves and exactly marginal deformations. We then move on to discuss some basic theory-specific consistency checks of our proposal that allow us to make contact with previous results in the literature. Afterwards, we describe our monopole vev RG flow formalism and explain how our formulae for the $(A_{N-1}, A_{N(n-1)-1})$ indices encode 4D ancestors of these flows. We then discuss the discrete symmetries of the index and the implications for the corresponding chiral algebras (in the sense of \cite{Beem:2013sza}). Before concluding, we speculate about the hypothetical exotic $\CN=2$ SCFTs alluded to above.

\section{AD theories from 6d (2,0) $A_{N-1}$ theories}\label{sec:Hitchin}
All of the $(A_{N-1}, A_{N(n-1)-1})$ and $(I_{N,N(n-1)}, R_Y)$ SCFTs discussed above are in class $\CS$. A theory, $\CT_{\CC}$, of class $\CS$ can be engineered by taking the six-dimensional $(2,0)$ theory with any ADE Lie algebra, 
$\mathfrak{g}$, and compactifying it on a (punctured) Riemann surface, $\CC$. In order to obtain a 4D theory with $\CN=2$ supersymmetry (SUSY), one performs a partial topological twist on $\CC$ (thereby breaking $SO(5)_R\to SO(2)_R\times SO(3)_R$) and imposes a BPS boundary condition at each puncture \cite{Gaiotto:2009hg,Gaiotto:2009we,Witten:1997sc}. One can understand these boundary conditions as corresponding to half-BPS co-dimension two defects that fill the four-dimensional spacetime and endow $\CT_{\CC}$ with flavor symmetries. 

In our earlier paper \cite{Buican:2015ina} we focused on the case $\mathfrak{g}=A_1$, while in this paper we generalize to the case $\mathfrak{g}=A_{N-1}$ for $N\ge2$. This theory supports a plethora of somewhat simpler to understand \lq\lq regular" defects \cite{Gaiotto:2009hg,Gaiotto:2009we,Gukov:2006jk,Chacaltana:2012zy} in addition to a bewildering array of so-called \lq\lq irregular" defects \cite{Witten:2007td,Gaiotto:2009hg,Bonelli:2011aa, Xie:2012hs,Wang:2015mra}. In all cases, we can describe the Coulomb branch physics and aspects of the flavor symmetries by considering a Higgs field, $\varphi$, of an $A_{N-1}$ Hitchin system living on $\CC$ (this is an ${(\rm End}V)$-valued meromorphic $(1,0)$ form on $\CC$, where $V$ is an $SU(N)$ bundle). Casimirs of $\varphi$ then correspond to vevs of $\CN=2$ chiral operators, $\langle\CO\rangle$, that parameterize the Coulomb branch of the 4D theory. In the case of a regular defect, $\varphi$ has a simple pole at the insertion point (which we refer to as a \lq\lq regular puncture"), and the corresponding flavor symmetry is a subgroup $G\subseteq SU(N)$ depending on the nature of the regular puncture.\footnote{Although it will not be particularly important in our discussion below, we note that it is possible for there to be additional flavor symmetries that are not manifest in this way of describing the physics.} If all the punctures are regular, the theory is typically superconformal.

On the other hand, in the case of an irregular defect, the $\varphi$ field has a pole of order $n+1$ (with $n>0$ referred to as the \lq\lq rank" of the defect) at the insertion point (which we refer to as an \lq\lq irregular puncture" on $\CC$). In the theories we study below, $n\in\mathbb{Z}_{\ge0}$, but there are more general possibilities. The flavor symmetry group induced by such a singularity will generically be $U(1)^{N-1}$ in the cases we consider below (but, again, there are more general possibilities). According to the discussion in \cite{Bonelli:2011aa,Xie:2012hs}, $\CT_{\CC}$ is an SCFT only if $\CC$ is a $\mathbb{CP}^1$ with either one irregular puncture and no regular punctures or with one irregular puncture and one regular puncture. In either of these two cases, $\CT_{\CC}$ will typically have non-integer scaling dimension $\CN=2$ chiral primaries and will therefore be an AD SCFT. We will encounter both situations below.

\subsection{Irregular singularities of type I}
In this section, we briefly describe the form of the Higgs field $\varphi$ near the irregular punctures we introduced above. The particular irregular singularities we are interested in are called \lq\lq type I" in the nomenclature of \cite{Xie:2012hs}. Placing such a singularity at the point $z=\infty$ on $\CC$ yields an expansion for $\varphi$ of the form
\begin{equation}\label{irregsing}
\varphi(z)=dz\left[M_1z^{n-1}+M_2z^{n-2}+\cdots+M_n+{M_{n+1}\over z}+\CO(z^{-2})\right]~,
\end{equation}
where the $M_i$ are arbitrary traceless $N\times N$ matrices. The singular terms above (i.e., those up to $\CO(z^{-2})$) encode the various relevant and exactly marginal deformations of the $\CT_{\CC}$ SCFT. In the case of the $(A_{N-1}, A_{N(n-1)-1})$ theories, $\varphi$ is regular for $z\ne\infty$. On the other hand, for the theories of type $(I_n, R_Y)$, $\varphi$ has an additional regular singularity at $z=0$. The corresponding simple pole encodes the additional flavor symmetry arising from the puncture at $z=0$.

\subsection{Seiberg-Witten curves of the AD theories and exactly marginal couplings}
Given the expansion for the Higgs field in \eqref{irregsing}, we can obtain the Seiberg-Witten curve for the $(A_{N-1},A_{N(n-1)-1})$ SCFT by considering the spectral curve (i.e., $\det\left(xdz-\varphi(z)\right)=0$) and using reparametrizations that do not affect the physics
\begin{align}\label{curve}
0 = x^{N} + a_2 x^{N-2}z^{2(n-1)} + a_3 x^{N-3}z^{3(n-1)} + \cdots + a_{N-1}xz^{(N-1)(n-1)} + z^{N(n-1)} + \cdots~,
\end{align}
where the final ellipses contain lower-dimensional terms.\footnote{Here the dimensions of the coordinates are $[x]={n-1\over n}$ and $[z]={1\over n}$. This statement follows from the fact that $\lambda=xdz$ is the 1-form and that therefore $[x]+[z]=1$.} The $a_i$ coefficients we have written explicitly in \eqref{curve} are dimensionless and give the $N-2$ exactly marginal couplings of the theory for $n>2$. Note that for $n=2$, there are only $N-3$ such couplings because one of the $a_i$ can be eliminated by a change of coordinates. There are also $N-1$ mass parameters corresponding to the deformations of the curve by $x^k z^{(n-1)(N-k)-n}$ for $k=0,\cdots,N-2$.

Hence, to summarize, for sufficiently large $N$ we have conformal manifolds of the following dimensions
\begin{eqnarray}\label{confmfld}
&&\dim_{\mathbb{C}}\left(\CM^{\rm conf}_{(A_{N-1},A_{N(n-1)-1})}\right)=N-2~, \ \ \ (n>2) ~,\cr&&\dim_{\mathbb{C}}\left(\CM^{\rm conf}_{(A_{N-1},A_{N-1})}\right)=N-3~.
\end{eqnarray}
Moreover, using methods described in \cite{Xie:2012hs}, it is straightforward to compute the corresponding conformal anomalies
\begin{eqnarray}\label{acAD}
a&=&{(N-1)(2N^4(n-1)^2+2N^3(n-1)^2-5N(n-1)-5)\over24(N(n-1)+1)}~, \cr c&=&{(N-1)(N(n-1)(N^2(1+N)(n-1)-2)-2)\over12(N(n-1)+1)}~.
\end{eqnarray}

The curves, spaces of marginal deformations, and anomalies for the theories with an additional regular singularity are more complicated, but can be found using similar methods. We will not discuss their detailed form here but instead refer the reader to \cite{Xie:2012hs,Xie:2013jc} for details.

\subsection{Reduction to three dimensions}
As we will see below, our formulas in \eqref{eq:wf1} and \eqref{eq:general} contain a wealth of information about RG flows between different theories of type $(A_{N-1}, A_{N(n-1)-1})$ via the residues of various poles in the Schur index. To check these predictions, we will find it useful to compare our results with RG flows for these same theories compactified on a circle (where the $(A_{N-1}, A_{N(n-1)-1})$ SCFTs flow to 3D $\CN=4$ SCFTs that are believed to be described by a 3D $\CN=4$ Lagrangian). The fundamental reason we can learn about 4D RG flows from 3D ones is that our procedure manifestly preserves eight Poincar\'e supercharges throughout and so non-perturbative superpotentials cannot be generated.

In fact, it will be simpler to study the 3D mirrors of the direct $S^1$ reductions of these theories \cite{Xie:2012hs,Xie:2013jc}. For the $(A_{N-1}, A_{N(n-1)-1})$ SCFT, the 3D mirror is given by a $U(1)^{N-1}$ quiver gauge theory with each $U(1)$ factor having $n-1$ fundamental hypermultiplets ($Q_{i,k}, \tilde Q_{i,k}^{\dagger}$ with $i\ne j$ labeling the $U(1)$ nodes) and $n-1$ bifundamental hypermultiplets ($Q_{i,j,k}, \tilde Q_{i,j,k}^{\dagger}$) between each pair of nodes (where we define $Q_{i,j,k}=Q_{j,i,k}, \tilde Q_{i,j,k}^{\dagger}=\tilde Q_{j,i,k}^{\dagger}$). The resulting $\CN=4$ superpotential reads
\begin{equation}\label{N4super}
W=\sum_{i=1}^{N-1}\sum_{k=1}^{n-1}\Phi_i\left(Q_{i,k}\tilde Q_{i,k}+q_{i,j}\sum_{j\ne i}Q_{i,j,k}\tilde Q_{i,j,k}\right)~,
\end{equation}
where $i,j=1,\cdots, N-1$, $k=1,\cdots,n-1$, and $q_{i,j}=-q_{j,i}=1$ for $i<j$.

The theories with an additional regular puncture have closely related 3D mirrors that have been proposed in \cite{Xie:2012hs,Xie:2013jc}. We will not discuss these latter dimensionally reduced theories further below, but our discussion of RG flows can be extended to these cases as well.

\section{The Schur index}
\label{sec:formula}
Before discussing our formulas in greater detail, we would like to briefly review the construction of the superconformal index. This quantity is a refined Witten index that counts operators in short representations of the superconformal group weighted by three superconformal fugacities ($p,q,t$) and arbitrarily many flavor fugacities ($x_i$). The counting is modulo short multiplets that can pair up to form long multiplets, thereby guaranteeing the invariance of the index under exactly marginal deformations (as long as the spectrum is discrete). The index is written with respect to some supercharge $\CQ$ (with $\left\{\CQ,\CQ^{\dagger}\right\}=\Delta$) and a mutually commuting set of charges as
\begin{equation}
\CI(p,q,t,x_i)={\rm Tr}_{\CH}(-1)^Fe^{-\beta\Delta}p^{j_1+j_2-r}q^{j_1-j_2-r}t^{R+r}\prod_i x_i^{f_i}~,
\end{equation}
where the trace is taken over the Hilbert space of local operators, $\CH$, $j_{1,2}$ are the $SO(4)$ spins, $R$ is the $SU(2)_R$ Cartan, $r$ is the $U(1)_R\subset U(1)_R\times SU(2)_R$ superconformal $R$-charge, and the $f_i$ are flavor charges. By standard arguments, only operators annihilated by $\CQ$ and $\CQ^{\dagger}$ contribute to the index (these states have $\Delta=E-2j_1-2R+r=0$ with $E$ the scaling dimension).

A simpler but nonetheless physically rich limit of the index, called the Schur index, is obtained by taking $t\to q$. It is straightforward to check that the $p$ dependence drops out and we are left with
\begin{equation}\label{Schurlim}
\CI(q,x_i)={\rm Tr}_{\CH}(-1)^Fe^{-\beta\Delta}q^{E-R}\prod_ix_i^{f_i}~.
\end{equation}
This limit of the index is intimately connected with 2D chiral algebras via the correspondence in \cite{Beem:2013sza}, with Coulomb branch physics \cite{Cecotti:2010fi,Buican:2015hsa,Cordova:2015nma,Fredrickson:2017yka} (even though the Coulomb branch operators themselves do not contribute to \eqref{Schurlim}), with Higgs branch physics (such operators, of type $\hat\CB_R$ in the nomenclature of \cite{Dolan:2002zh} (see also \cite{Dobrev:1985qv}), contribute directly to \eqref{Schurlim}),\footnote{We can think of Higgs branch physics as contributing perturbatively in $q$ to \eqref{Schurlim} while Coulomb branch physics contributes only non-perturbatively in $q$ \cite{Buican:2015hsa}.}, $S^3$ partition functions for dimensionally reduced theories \cite{Buican:2015hsa,Gadde:2011ia,Nishioka:2011dq}, and, crucially for us below, with $q$-deformed Yang-Mills theory in the class $\CS$ context \cite{Gadde:2011ik}. In the next subsection, we will motivate our proposals \eqref{eq:general} and \eqref{2puncture} for this limit of the index.

\subsection{Motivating our generalization}
\label{subsec:motivation}

To explain our proposal in \eqref{eq:general}, it is useful to recall the corresponding result in the case of $SU(N)$ $q$-deformed Yang-Mills theory on a Riemann surface, $\CC$, of genus $g$ with $m$ regular punctures (defined by Young diagrams, $Y_i$, with $i=1,\cdots,m$) \cite{Gadde:2011ik}. Indeed, the authors of \cite{Gadde:2011ik} argued that the Schur index in this case can be computed (up to an overall prefactor) as an $m$-point correlator in the zero-area limit of $q$-deformed Yang-Mills theory
\begin{equation}\label{regex}
\CI_{\CT_{\CC}}(q;{\bf x})=\sum_RC_R(q)^{2-2g-m}\prod_{k=1}^mf_R^{Y_k}(q;{\bf x}_k)~,
\end{equation}
where $C_R(q)$ is defined as in \eqref{topological}, and ${\bf x}_k\equiv(x_{1,k},\cdots, x_{N-1,k})$ are $SU(N)$ fugacities. The $R$ dummy variables in \eqref{regex} correspond to irreducible representations of $SU(N)$ labeling intermediate states in the TFT correlator, and $f_R^{Y_k}$ is the inner product of the state $|{\bf x}_k,Y_k\rangle$ corresponding to the holonomy around the $k$-th regular puncture with the state corresponding to the representation $R$, i.e.
\begin{equation}
\langle R|{\bf x}, Y_k\rangle=f_R^{Y_k}(q;{\bf x}_k)~.
\end{equation}
The topological nature of the index in \eqref{regex} is reflected in the fact that it is independent of the ordering of the punctures.

Therefore, in order to compute the index for the $(A_{N-1}, A_{N(n-1)-1})$ theory from this perspective, the main question is  to fix the irregular wave function, $\tilde f^{(n)}_R(q;{\bf x})$. Indeed, the factor of $C_R(q)$ in \eqref{eq:general} follows from the discussion in the regular puncture case by noting that $\CC=\mathbb{CP}^1$ with a single puncture (and so $2-2g-m=1$).\footnote{Similarly, there is no factor of $C_R(q)$ in \eqref{2puncture} since $2-2g-m=0$.}

In our earlier paper \cite{Buican:2015ina} we provided strong evidence (confirmed using other techniques \cite{Cordova:2015nma,Maruyoshi:2016aim}) that, in the $SU(2)$ case, the index is given by \eqref{eq:general} with 
\begin{equation}\label{su2wfn}
\tilde f^{(n)}_R(q;x)=\prod_{k=1}^{\infty}\left({1\over1-q^k}\right)q^{nC_2(R)}{\rm Tr}_R\left[x^{2J_3}q^{-n(J_3)^2}\right]~,
\end{equation}
where $J_3=\pm{1\over2}$ for the fundamental representation. One important aspect of this formula is that it is manifestly invariant under the $SU(2)$ Weyl group (in this case $S_2\simeq Z_2$). This invariance is natural since the Hitchin system description of our theory has $S_N$ invariance.

To emphasize this invariance and also to see how to generalize \eqref{su2wfn} in an $SU(N)$ Weyl-invariant way, we can re-write our expression as
\begin{eqnarray}\label{rewrite}
\tilde{f}^{(n)}_R(q;x)&=&\prod_{k=1}^\infty \left(\frac{1}{1-q^k}\right)q^{nC_2(R)}\text{Tr}_R\left[q^{-\frac{n}{2}h_1\left(\frac{1}{2}\right)h_1}x^{h_1}\right]=\cr&=& \prod_{k=1}^\infty\left(\frac{1}{1-q^k}\right)q^{nC_2(R)}\text{Tr}_R\left[q^{-\frac{n}{2}F^{11}h_1h_1}x_1^{h_1}\right]~,
\end{eqnarray}
where the various quantities appearing are written in terms of the Chevalley basis $e_1,\, f_1,\,h_1$ of $SU(2),$\footnote{Recall again that the Chevalley basis satisfies $[h_i,\,e_j] = A_{ji}e_j,\, [h_i,\,f_j] = -A_{ji}f_j$ and $[e_i,\,f_j] = \delta_{ij}h_j$, where $A_{ij}$ is the Cartan matrix. In the case of $SU(2)$, we can identify $J_3 = \frac{1}{2}h_1$.} and $F^{11}={1\over2}$ is the (in this case $1\times 1$) inverse Cartan matrix appearing in \eqref{Cartinv} (with $N=2$). Weyl invariance follows from the fact that any highest weight representation, $R$, is spanned by states $|\lambda\rangle$ such that $h_i|\lambda\rangle = \lambda_i |\lambda\rangle$, and $F^{ij}\lambda_i\lambda_j = (\lambda,\lambda)$ is Weyl invariant.

It is now clear how to generalize the $R$-dependent part of in \eqref{rewrite} for general $N$ in a minimal way that respects the $SU(N)$ Weyl invariance (recall that ${\rm Weyl}[SU(N)]\simeq S_N)$ of $q$-deformed Yang-Mills theory
\begin{equation}\label{weyltr}
q^{nC_2(R)}\text{Tr}_R\left[q^{-\frac{n}{2}F^{11}h_1h_1}x_1^{h_1}\right]\to q^{nC_2(R)}\text{Tr}_R\left[q^{-\frac{n}{2}F^{ij}h_ih_j}\prod_{i=1}^{N-1}(x_1\cdots x_i)^{h_i}\right]~.
\end{equation}

Generalizing the first factor in \eqref{su2wfn} is also straightforward. In \cite{Buican:2015ina}, we argued that, for $N=2$, \eqref{su2wfn} was the natural generalization of the plethystic exponential in the regular puncture wavefunction \eqref{Regular} to the irregular wave function case, where the global symmetry is generically reduced to $U(1)$. For $N\ge2$, the global symmetry is generically $U(1)^{N-1}$, and so a straightforward generalization gives
\begin{equation}\label{cartans}\prod_{k=1}^{\infty}\left({1\over1-q^k}\right)\to\prod_{k=1}^{\infty}\left({1\over1-q^k}\right)^{N-1}~.
\end{equation}
This latter expression has the added benefit that, when combined with $C_R(q)$, it gives the expected asymptotic Cardy-like behavior in the limit $q\to1$ (we will make this statement more precise below).

Finally, our expression has the important property that
\begin{eqnarray}\label{bound}
\tilde{f}_R^{(n)}(q; {\bf x})  =  O\left( q^{n(\lambda, \rho)} \right)~, \ \ \ {\rm dim}_q R  =  O\left( q^{-(\lambda, \rho)} \right)~,
\end{eqnarray}
where $\lambda$ is the Dynkin label of the highest weight of the representation, $R$, of $SU(N)$, $\rho$ is the Weyl vector of $SU(N)$, and the inner product is the standard one
\begin{equation}
(\lambda,\rho)\equiv\sum_{i,j}\lambda_iF^{ij}\rho_j=\sum_{i,j}\lambda_iF^{ij}~,
\end{equation}
where we have used the fact that $\rho=(1,\cdots,1)$. One immediate consequence of \eqref{bound} is that our expressions for the indices have only positive powers of $q$ (as required by unitarity). Combined with the fact that $C_R(q)$ has only positive contributions in $q$ (in the sense of the coefficients), we have that flavor contributions to the index obey
\begin{equation}\label{flavor}
{\rm Flavor} = \CO\left(q^{(n-1)(\lambda,\rho)}\right)~.
\end{equation}
We will be able to use \eqref{flavor} to argue that our indices capture the correct symmetry structure of the $(A_{N-1}, A_{N(n-1)-1})$ SCFTs and to explain how baryon contributions to the index arise from $SU(N)$ representations of $q$-deformed Yang-Mills theory.

\subsection{Some general properties}
On general grounds, the $O(q)$ terms in the Schur index can only come from flavor symmetry moment maps.\footnote{Via the correspondence of \cite{Beem:2013sza}, which maps Schur indices in 4D to vacuum characters of chiral algebras in 2D and 4D flavor symmetry moment maps to 2D affine Kac-Moody (AKM) currents, this statement corresponds to the fact that any $O(q)$ term in the vacuum character of a chiral algebra is an AKM current (here we normalize the character so that its expansion starts with a \lq\lq 1").} In the $(A_{N-1}, A_{N(n-1)-1})$ theories, we have a flavor symmetry of rank $N-1$. In the case $n>3$, this symmetry is simply $U(1)^{N-1}$. For $n=3$ and $N=2$, the symmetry is enhanced to $SU(2)$, while it remains $U(1)^{N-1}$ for $N>2$. Finally, for $n=2$ and $N=2$ we have $SU(2)$ flavor symmetry, for $N=3$ we have $SU(3)$ flavor symmetry, and for $N>3$ we again have $U(1)^{N-1}$ flavor symmetry.

Using the result in \eqref{flavor}, we can easily argue for this flavor structure. First, suppose $n>3$. In this case, we expect $U(1)^{N-1}$ flavor symmetry and hence
\begin{equation}\label{abflavor}
\CI_{A_{(N-1}, A_{N(n-1)-1})}=1+(N-1)q+\cdots~, \ \ \ n>3~, 
\end{equation}
since the adjoint of $U(1)^{N-1}$ is $N-1$ singlet representations. Indeed, from \eqref{flavor}, we see that the power of $q$ at which flavor comes in is
\begin{equation}\label{flbound}
(n-1)(\lambda,\rho)\ge{(n-1)(N-1)\over2}~.
\end{equation}
For $n>3$, we see that the flavor contributions come in at order $q^{3\over2}$ and higher. Therefore, we can only have singlets at $O(q)$. In this case, it is easy to see that the only contributions at $O(q)$ come from the $(1-q)^{-(N-1)}=1+(N-1)q+\cdots$ factor in \eqref{cartans}.

Next, consider the case $n=3$. Clearly, if $N>2$, then by \eqref{flbound} flavor contributions come in at order $q^{2}$ and higher, and we can again argue that the index takes the form in \eqref{abflavor}. On the other hand, if $N=2$, then there will be contributions at $O(q)$. These contributions are from $SU(2)$ moment maps corresponding to raising and lowering operators of $SU(2)$ and were described in our paper \cite{Buican:2015ina}.

Finally, consider setting $n=2$. In this case, if $N>3$, then flavor contributions come in at order $q^{3\over2}$, and we are back to the analysis presented in the $n>3$ case. If $N=3$, then there are flavor contributions at order $q$ corresponding to the flavor symmetry enhancement $U(1)^2\to SU(3)$ of the $(A_2, A_2)\simeq(A_1, D_4)$ theory. The $(A_1, D_4)$ index was described in \cite{Buican:2015ina} and corresponds to the vacuum partition of $\widehat{su(3)}_{-{3\over2}}$. Below we will describe the same theory from the dual $(A_2, A_2)$ perspective and explicitly check this flavor symmetry enhancement. If $N=2$, then we are in the case of the free hypermultiplet, $(A_1, A_1)$. This theory again has flavor contributions at order $q$ reflecting the $U(1)\to SU(2)$ flavor symmetry enhancement (this case was covered in \cite{Buican:2015ina}).

The result in \eqref{flavor} also allows us to describe how baryons arise in the index of the $(A_{N-1}, A_{N(n-1)-1})$ SCFT. Indeed, these theories have baryons (and also monopoles in the mirrors of the $S^1$ reductions that we will study further below)
\begin{equation}\label{baryoncont}
(E-R)|_{\rm Baryons} =E^{3d}_{\rm Monopoles}= {(n-1)\nu(N-\nu)\over2}~, \ \ \ \nu=1,\cdots,N-1~.
\end{equation}
Using our formula in \eqref{flavor}, we can reproduce these baryonic contributions from $SU(N)$ representations with Dynkin labels
\begin{equation}\label{baryonreps}
\lambda_i = \delta_{i,\nu}~,
\end{equation}
thus furnishing an interesting algebraic description of these operators and the corresponding moduli (sub)-spaces they parameterize.

Let us also note that if we have a conformal manifold with $n=2$ and complex dimension $N-3>0$ as in \eqref{confmfld}, then the flavor symmetry is $U(1)^{N-1}$ and there are at least three moment maps at $O(q)$. These moment maps are related to three AKM currents via the correspondence in \cite{Beem:2013sza} and hence we find that the resulting chiral algebra has at least three generators in accord with the general bounds of \cite{Buican:2016arp}. If we have a conformal manifold with $n>2$ and complex dimension $N-2>0$, then the flavor symmetry is at least $U(1)^2$. However, we must also have baryons from \eqref{baryoncont}. These are generators of the Hall-Littlewood ring \cite{Gadde:2011uv} and hence are also generators of the corresponding chiral algebra \cite{Beem:2013sza} and so the bounds of \cite{Buican:2016arp} are again obeyed.

Finally, we close by noting that this discussion on the irregular singularities and the flavor symmetry guarantees that the theories with an additional regular singularity as in \eqref{2puncture} have the correct flavor symmetry dependence at $O(q)$ as well.

\section{Consistency checks}
\label{sec:consistency}
In the first part of this section, we perform various non-trivial consistency checks of our generalization \eqref{eq:general} by matching it onto previously known indices that can be constructed by duality with $A_1$ Hitchin systems or by conformally gauging isolated theories constructed from $A_1$ Hitchin systems for small $N$ and $n$.

We then describe the proposed Lagrangian mirrors of the $S^1$ reductions of the $(A_{N-1}, A_{N(n-1)-1})$ SCFTs and certain RG flows that interpolate between such theories. These flows involve giving vevs to monopole operators, and we will present a simple formalism to capture the resulting physics. We then compare our results with the parent RG flows in 4D encoded in our indices via the procedure described in \cite{Gaiotto:2012xa} (and already exploited in the $A_1$ irregular singularity case in \cite{Buican:2015ina}). This gives a strong check of our proposal for all $N, n\ge2$ and also of the Lagrangians in \cite{Xie:2012hs,Xie:2013jc}. Finally, we conclude with a non-perturbative-in-$q$ check of the expected Cardy-like behavior of these indices \cite{DiPietro:2014bca,Ardehali:2015bla,Buican:2015ina,DiPietro:2016ond}.

\subsection{Lower rank checks}
\label{subsec:lower}
The checks that follow, while highly non-trivial, are only for small $N, n$. In the next subsection, we will give checks for all $N, n\ge2$. Since we have already seen that our formula for general $N$ is consistent with our earlier work on $N=2$ in \cite{Buican:2015ina}, here we focus on the case of $N>2$.

\subsubsection{The $(A_2,A_2)$ index}
The $(A_2, A_2)$ theory is dual to the $(A_1, D_4)$ theory \cite{Cecotti:2010fi}. In \cite{Buican:2015ina} we gave a closed-form expression for the Schur index of this latter theory (it is just the vacuum character of $\widehat{su(3)}_{-{3\over2}}$).

Given our generalization, we see that the $(A_2, A_2)$ index (and hence the vacuum character of $\widehat{su(3)}_{-{3\over2}}$) must also be expressible as
\begin{align}
\mathcal{I}_{(A_2,A_2)}(q;x,y) &= \sum_{R}C_R\, \tilde{f}_R^{(2)}(q;x,y)~,
\end{align}
where $R$ runs over irreducible $SU(3)$ representations, and $C_R$ is given by \eqref{topological} (with $N=3$)
\begin{align}
C_R = \frac{(1-q)^2(1-q^2)}{(q;q)_\infty^2}\chi_R^{su(3)}(q,1,q^{-1})~.
\label{eq:A2-C_R}
\end{align}
In this case, our proposed wave function is
\begin{align}
\tilde{f}_{R}^{(2)}(q;x,y) = \left(\prod_{k=1}^\infty \frac{1}{1-q^k}\right)^2 q^{2C_2(R)} \text{Tr}_{R}\left[q^{-F^{ij}h_ih_j}x_1^{h_1}(x_1x_2)^{h_2}\right]~.
\end{align}
We have checked that the above formula reproduces the $(A_1,D_4)\simeq (A_2,A_2)$ index that we described in \cite{Buican:2015ina} correctly up to high order in $q$.

\subsubsection{$(A_3,A_3)$ index}

For the $(A_3,A_3)$ theory, our above discussion implies that
\begin{align}
 \mathcal{I}_{(A_3,A_3)}(q;x_1,x_2,x_3) = \sum_{R}C_R\, \tilde{f}_R^{(2)}(q;x_1,x_2,x_3)~,
\label{eq:A3A3}
\end{align}
where
\begin{align}
C_R 
=  \frac{(1-q)^3(1-q^2)^2(1-q^3)}{(q;q)_\infty^3}\chi_{R}^{su(4)}(q^{\frac{3}{2}},q^{\frac{1}{2}},q^{-\frac{1}{2}},q^{-\frac{3}{2}})~.
\end{align}
The wave function corresponding to the irregular puncture is given by
\begin{align}
 \tilde{f}^{(2)}_R(q;x_1,x_2,x_3) &= \left(\prod_{k=1}^\infty\frac{1}{1-q^k}\right)^3 q^{2C_2(R)}\text{Tr}_R\left[q^{-F^{ij}h_ih_j}x_1^{h_1} (x_1x_2)^{h_2} (x_1x_2x_3)^{h_3}\right]~,
\end{align}
where $R$ runs over irreducible representations of $SU(4)$.

\begin{figure}
\begin{center}
\vskip .4cm
\begin{tikzpicture}[place/.style={circle,draw=blue!50,fill=blue!20,thick,inner sep=0pt,minimum size=7mm},transition/.style={rectangle,draw=black!50,fill=black!20,thick,inner sep=0pt,minimum size=6mm},transition2/.style={rectangle,draw=black!50,fill=red!20,thick,inner sep=0pt,minimum size=8mm},auto]

\node[transition2] (1) at (-2,0) {\,$(A_1,D_4)$\,};
\node[place] (2) at (0,0) [shape=circle] {$2$} edge [-] node[auto]{} (1);
\node[transition2] (3) at (2,0)  {\;$(A_1,D_4)$\;} edge [-] node[auto]{} (2);
\node[transition] (9) at (0,-1.4) {$\;1\;$} edge[-] (2);
\end{tikzpicture}
\caption{A quiver diagram describing a weak coupling limit of the $(A_3,A_3)$ theory.}
\label{fig:A3A3}
\end{center}
\end{figure}
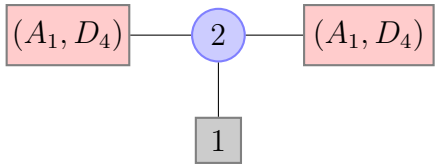

Our expression \eqref{eq:A3A3} for the index can be expanded in powers of $q$ as follows
\begin{align}
\mathcal{I}_{(A_3,A_3)}(q;x_1,x_2,x_3) =& 1 + 3q + (a+1/a)(b+1/b+c+1/c)q^{\frac{3}{2}}
\nonumber\\
& + \left[10 + a^2+1/a^2+(b+1/b)(c+1/c)\right]q^2
\nonumber\\
& + 4(a+1/a)(b+1/b+c+1/c)q^{\frac{5}{2}} + \cdots~,
\end{align}
where we reparameterize $x_{1,2,3}$ by $x_1 \equiv a b,\, x_2 = c/a$, and $x_3=a/b$.
We have checked that the above expression coincides with the $(A_3,A_3)$ index evaluated in \cite{Buican:2015ina} by gauging a diagonal $SU(2)$ subgroup of the flavor symmetry of two $(A_1, D_4)$ theories and two $(A_1, A_1)$ theories (our expression here is considerably simpler since there is no integration over a gauge group). In terms of the quiver description in figure \ref{fig:A3A3}, the fugacity $a$ corresponds to the $U(1)$ flavor symmetry acting on the hyper multiplets while $b$ and $c$ are fugacities associated with the two $(A_1,D_4)$ theories.

The S-duality transformation (see \cite{Buican:2014hfa,DelZotto:2015rca,Cecotti:2015hca} for a discussion of this duality) $(a,b,c)\to (\sqrt{b/c},\,a\sqrt{bc},\,\sqrt{bc}/a)$ corresponds to 
\begin{align}
x_3 \to \frac{1}{x_1x_2x_3}~,\ \ \ \text{with}\ \  x_1~,x_2 \ \ \text{fixed}~.
\label{eq:S-dual}
\end{align}
This transformation is identical to the Weyl reflection associated with $\alpha_3$.\footnote{Indeed, the above transformation implies $\lambda_k \to \lambda_k- \langle \alpha_k,\alpha_3\rangle\lambda_3$, or, equivalently, $\lambda \to \lambda - \langle \lambda,\alpha_3\rangle \alpha_3$.} Since \eqref{eq:A3A3} is invariant under the action of the Weyl group of $A_3$, our conjecture \eqref{eq:A3A3} is manifestly invariant under the S-duality transformation.\footnote{Recall from the discussion above that the Weyl invariance follows from the fact that any highest weight representation $R$ is spanned by states $|\lambda\rangle$ such that $h_i|\lambda\rangle = \lambda_i |\lambda\rangle$, and $F^{ij}\lambda_i\lambda_j = (\lambda,\lambda)$ is Weyl invariant.} Note also that the other generators of the Weyl group exchange $x_1$ and $x_2$, or $x_2$ and $x_3$. In terms of the quiver description shown in figure \ref{fig:A3A3}, these two exchanges correspond to the combination of \eqref{eq:S-dual} and a symmetry manifest in the weak coupling description shown in Fig~.\ref{fig:A3A3}.

\subsubsection{$(A_2,A_5)$ theory}
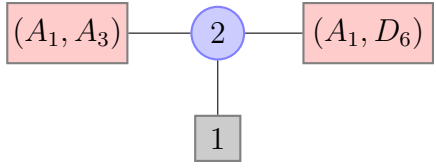
\begin{figure}
\begin{center}
\vskip .4cm
\begin{tikzpicture}[place/.style={circle,draw=blue!50,fill=blue!20,thick,inner sep=0pt,minimum size=7mm},transition/.style={rectangle,draw=black!50,fill=black!20,thick,inner sep=0pt,minimum size=6mm},transition2/.style={rectangle,draw=black!50,fill=red!20,thick,inner sep=0pt,minimum size=8mm},auto]

\node[transition2] (1) at (-2,0) {\,$(A_1,A_3)$\,};
\node[place] (2) at (0,0) [shape=circle] {$2$} edge [-] node[auto]{} (1);
\node[transition2] (3) at (2,0)  {\;$(A_1,D_6)$\;} edge [-] node[auto]{} (2);
\node[transition] (9) at (0,-1.4) {$\;1\;$} edge[-] (2);
\end{tikzpicture}
\caption{A quiver diagram describing a weak coupling limit of the $(A_2,A_5)$ theory.}
\label{fig:A2A5}
\end{center}
\end{figure}

A simple $S$-duality for the $(A_2, A_5)$ theory was worked out in \cite{DelZotto:2015rca} and was studied at the level of the Macdonald index in \cite{Buican:2015tda}. This theory has a quiver description given in Fig. \ref{fig:A2A5} which enabled us to compute its Macdonald index in \cite{Buican:2015tda} as an integral over the diagonal $SU(2)$ gauge group. On the other hand, from our expressions above, we see that the Schur index of this theory should be expressible in terms of an expansion over $SU(3)$ characters
\begin{align}
\mathcal{I}_{(A_2,A_5)} (q;x_1,x_2) = \sum_{R}C_R\, \widetilde{f}_{R}^{(3)}(q;x_1,x_2)~,
\end{align}
where $C_R$ is given by \eqref{eq:A2-C_R}, and the irregular wave function is
\begin{align}
\widetilde{f}_{R}^{(3)} =& \left(\prod_{k=1}^\infty\frac{1}{1-q^k}\right)^2 q^{3C_2(R)}\text{Tr}_R\left[q^{-\frac{3}{2}F^{ij}h_ih_j} x_1^{h_1}(x_1x_2)^{h_2}\right]~.
\end{align}
By an explicit computation, we obtain
\begin{align}
\mathcal{I}_{(A_2,A_5)}(q;x_1,x_2) =& 1 + 2q + \left(6+c^2 + bc + \frac{b}{c} + \frac{c}{b} + \frac{1}{bc} + \frac{1}{c^2}\right)q^2\nonumber\\
& \qquad + \left(14 + 3c^2 + 3bc + \frac{3c}{b} + \frac{3b}{c} + \frac{3}{bc} + \frac{3}{c^2}\right)q^3 +  \cdots~,
\label{eq:A2A5}
\end{align}
where the flavor fugacities $x_{1,2}$ are rewritten in terms of $b \equiv x_1\sqrt{x_2}$ and $c\equiv 1/\sqrt{x_2}$.
The above expression coincides with the index of the quiver theory shown in figure \ref{fig:A2A5}, where $b$ is the flavor fugacity for the $U(1)\supset  U(1)\times SU(2)$ flavor symmetry of the $(A_1,D_6)$ theory while $c$ is the fugacity for $U(1)$ flavor symmetry acting on the hyper multiplets. 

The S-duality transformation is, as found in \cite{Buican:2015tda}, given by $(b,c)\to (\sqrt{c^3/b},\, 1/\sqrt{bc})$. Indeed, the index \eqref{eq:A2A5} is invariant under this transformation. Up to the symmetry $b\to b^{-1}$, which is manifest in the weak coupling description shown in Fig.~\ref{fig:A2A5}, this S-dual transformation is equivalent to 
\begin{align}
x_2 \to \frac{1}{x_1x_2}~,\ \ \  x_1 \text{  : fixed}~,
\end{align}
which corresponds to the Weyl transformation associated with $\alpha_2$.

\subsection{RG flows}
In this section, we give a general non-perturbative check of our proposal in \eqref{eq:general} for all $N, n\ge2$. The RG flows we will discuss can be studied both in 4D and 3D and are interpolations between the following SCFTs in our class of theories
\begin{equation}\label{genflow}
(A_{N-1}, A_{N(n-1)-1})\to (A_{\nu-1},A_{\nu(n-1)-1})\oplus(A_{N-\nu-1},A_{(N-\nu)(n-1)-1})\oplus (A_1, A_1)~, \ \ \ 1<\nu< N-1~,
\end{equation}
as well as
\begin{equation}\label{origflow}
(A_{N-1}, A_{N(n-1)-1})\to(A_{N-2}, A_{(N-1)(n-1)-1})\oplus(A_{1}, A_{1})~.
\end{equation}
In the 4D description, these RG flows are triggered by turning on baryonic vevs.\footnote{The decoupled $(A_1, A_1)$ factors on the right-hand side (RHS) of \eqref{genflow} and \eqref{origflow} contain the axion-dilaton of spontaneous conformal breaking and their $\CN=2$ partners.} On general grounds \cite{Intriligator:1996ex}, in the mirror description of the $S^1$ reductions of these theories, the RG flow should be triggered by turning on vevs for monopole operators.

The consistency of our RG flow picture in both 4D and 3D is a highly non-trivial check of our proposal in \eqref{eq:general} as well as for the 3D mirrors of the dimensional reductions of the $(A_{N-1}, A_{N(n-1)-1})$ SCFTs in \cite{Xie:2012hs,Xie:2013jc}.\footnote{In the case of the $(A_1, A_{2n-3})$ and $(A_1, D_{2n})$ theories, we have additional checks of the 3D mirrors arising from the $S^3$ partition function computations in \cite{Buican:2015hsa} and the Higgs branch Hilbert series analysis in \cite{DelZotto:2014kka} (the Hilbert series analysis counts a proper subset of the Schur operators but also applies to the other $(A_N, A_M)$ theories).}

\subsubsection{The 3D picture and the mirror RG flow}\label{3DRG}
Around \eqref{N4super}, we briefly described the mirrors of the $S^1$ reductions of the $(A_{N-1}, A_{N(n-1)-1})$ SCFTs \cite{Xie:2013jc}: they are 3D $\CN=4$ SCFTs that arise from gauge-coupling RG flows of Lagrangian $U(1)^{N-1}$ quiver gauge theories with each $U(1)$ factor having $n-1$ fundamental hypermultiplets ($Q_{i,k}, \tilde Q_{i,k}^{\dagger}$ with $i\ne j$ labeling the $U(1)$ nodes) and $n-1$ bifundamental hypermultiplets ($Q_{i,j,k}, \tilde Q_{i,j,k}^{\dagger}$) between each pair of nodes (where we define $Q_{i,j,k}=Q_{j,i,k}, \tilde Q_{i,j,k}^{\dagger}=\tilde Q_{j,i,k}^{\dagger}$). The resulting $\CN=4$ superpotential was described in \eqref{N4super} and is reproduced below for ease of reference
\begin{equation}\label{N4super2}
W=\sum_{i=1}^{N-1}\sum_{k=1}^{n-1}\Phi_i\left(Q_{i,k}\tilde Q_{i,k}+q_{i,j}\sum_{j\ne i}Q_{i,j,k}\tilde Q_{i,j,k}\right)~,
\end{equation}
where $i,j=1,\cdots, N-1$, $k=1,\cdots,n-1$, and $q_{i,j}=-q_{j,i}=1$ for $i<j$.

The particular 3D flows we will study are triggered by turning on vevs for monopole operators in the above theories (so that the 4D flows correspond to flows induced by turning on vevs for baryons). These operators are superconformal primaries (at the IR endpoint of the flows described by the Lagrangians in \eqref{N4super2}) that have scaling dimension \cite{Gaiotto:2008ak}
\begin{equation}
E^{3d}_{\rm Monopoles} = j_L ={n-1\over2}\left(\sum_{i=1}^{N-1}|a_i|+\sum_{i<j}|a_i-a_j|\right) ~,
\end{equation}
where $j_L$ is the spin under the $SU(2)_L\subset SO(4)_R$ $R$-symmetry that acts on the (mirror) Coulomb branch, and $a_i\in\mathbb{Z}$ is the charge of the monopole operator under the $i$th $U(1)$ topological symmetry (there are $N-1$ such global symmetries corresponding to each $U(1)\subset U(1)^{N-1}$ gauge factor). By mirror symmetry, these operators map to baryons that have $E^{3d}_{\rm Baryons}=j_R$, where $j_R$ is the spin under the $SU(2)_R\subset SO(4)_R$ $R$-symmetry factor that acts on the Higgs branch.\footnote{Therefore, these baryons descend from 4D baryons of $SU(2)_R$ spin $j_R$ and scaling dimension $E_{\rm Baryons}=2j_R$.} The topological charges map to charges under the baryonic symmetries of the $(A_{N-1}, A_{N(n-1)-1})$ SCFTs in 4D.

Since monopoles do not appear in the superpotential \eqref{N4super2}, describing the flows resulting from turning on their vevs is non-trivial.\footnote{Although such an analysis can in principle be carried using standard matter operators built out of fields appearing in the superpotential of the mirror theory to \eqref{N4super2} (using the techniques in \cite{deBoer:1996ck}), this avenue becomes very tedious for the theories we study once $N$ and $n$ become sufficiently large.} Let us first study the somewhat simpler flow arising from the mirror of the $S^1$ reduction of \eqref{origflow}
\begin{equation}\label{origflow2}
(A_{N-1}, A_{N(n-1)-1})\to(A_{N-2}, A_{(N-1)(n-1)-1})\oplus(A_{1}, A_{1})~.
\end{equation}
Note that the $N=2$ case was already studied in \cite{Buican:2015ina} (the first factor on the RHS disappears in this case) where the flow was analyzed in the direct $S^1$ reduction. Here we give a simple description of this flow in the mirror theory for $N\ge2$. In this case, we claim that the flow in \eqref{origflow2} can be triggered by turning on the following vev
\begin{equation}\label{monovevsymm}
\langle\mathcal{O}_{1,\cdots,1}\rangle\ne0~,
\end{equation}
where $\CO_{1,\cdots,1}$ is the highest $SU(2)_L$ weight component of the monopole primary with $a_1=a_2=\cdots=a_{N-1}=1$ and
\begin{equation}
E^{3d}(\CO_{1,\cdots,1})={(N-1)(n-1)\over2}=I_3^L(\CO_{1,\cdots,1})~,
\end{equation}
where the RHS denotes the $SU(2)_L$ weight of the operator.\footnote{This flow can also be triggered by turning on a similar vev for the $j$th monopole with $a_i=\delta_{i,j}$ and $j=1,\cdots,N-1$. Together with the monopole we explicitly consider in the text and the same monopoles but with $a_i\to-a_i$, the 4D baryonic ancestors of these operators contribute to the Schur index as a fundamental plus an anti-fundamental of $SU(N)$ $q$-deformed Yang-Mills theory.}

To analyze the flow\footnote{Although it will not be important for us below, we note that the necessary RG analysis can be more subtle in theories with accidental IR superconformal $R$ symmetries (i.e., theories that are often referred to as \lq\lq bad" and \lq\lq ugly").} resulting from \eqref{monovevsymm}, we note that, by Goldstone's theorem, {\bf(i)} there will be a Goldstone Boson for the broken overall topological symmetry in the IR, and {\bf(ii)} this particle must be coupled irrelevantly to the rest of the theory. Moreover, at long distances, the corresponding topological current must have the form $j_{\mu}=\partial_{\mu}\phi$, where $\phi$ is the Goldstone Boson. In the case of the topological symmetry, we know $j_{\mu}=\left(\star F\right)_{\mu}$ and so the Goldstone boson is just the scalar dual to the photon (see \cite{Gaiotto:2014kfa} for related discussions).

Therefore, we see that the overall $U(1)$ vector multiplet must decouple in the IR. In particular, it cannot couple to anything charged under the corresponding gauge symmetry. This fact implies that, for each $U(1)_i$ node ($i=1,\cdots, N-1$) all $n-1$ fundamentals get a mass. On the other hand, the bi-fundamentals are all neutral under the overall $U(1)$ gauge symmetry and are therefore unaffected by the VEV in \eqref{monovevsymm}. Once we remove the overall $U(1)$, the remaining theory is precisely the $U(1)^{N-2}$ gauge theory that describes the mirror of the $S^1$ reduction of the $(A_{N-2}, A_{(N-1)(n-1)-1})$ theory. Combined with the decoupled free vector (which in 4D becomes a decoupled $(A_1, A_1)$ theory), we find the RG flow in \eqref{origflow}.

We claim that the more general RG flows in \eqref{genflow} can be analyzed in a similar spirit by turning on a VEV for any one of the monopoles with charges
\begin{equation}
a_{i_1}=a_{i_2}=\cdots=a_{i_{\nu}}=1~, \ \ \ i_{a}\ne i_b\ \forall a\ne b~,
\end{equation}
and all other topological quantum numbers vanishing.\footnote{There are ${(N-1)!\over(N-\nu-1)!\nu!}$ such operators. Combined with another set of ${(N-1)!\over(\nu-1)!(N-\nu)!}$ operators with $a_{j_1}=a_{j_2}=\cdots=a_{j_{N-\nu}}=1$ and $j_{a}\ne j_b$ $\forall a\ne b$ (and all other topological quantum numbers vanishing) and similar operators with $a_i\to-a_i$, we get precisely the number of operators in a $\nu$-index antisymmetric representation of $SU(N)$ and its conjugate, i.e., an $N-\nu$-index antisymmetric representation (unless $N$ is even and $\nu={N\over2}$; in this case we simply get a single ${N\over2}$-index antisymmetric combination). All in all, we see that the monopole operators are in one-to-one correspondence with the $\nu$-index anti symmetric representations of $SU(N)$ for $\nu=1,\cdots,N-1$ (i.e., the representations discussed around \eqref{baryonreps}).} This monopole operator has scaling dimension 
\begin{equation}
E^{3d}(\CO_{1,\cdots,1})={(N-\nu)\nu(n-1)\over2}=I_3^L(\CO_{1,\cdots,1})~.
\end{equation}
Moreover, reasoning similar to the one above with $\nu=N-1$ (now we take the overall $U(1)$ to be $U(1)_{a_{i_1}}+\cdots+ U(1)_{a_{i_{\nu}}}$) suggests that we have the following RG flow when turning on a VEV for this operator
\begin{equation}
(A_{N-1}, A_{N(n-1)-1})\to (A_{\nu-1},A_{\nu(n-1)-1})\oplus(A_{N-\nu-1},A_{(N-\nu)(n-1)-1})\oplus (A_1, A_1)~,
\end{equation}
where $1<\nu<N-1$.

Next we will describe the 3D Coulomb branch flows of this sub-section as Higgs branch flows in the parent 4D theory. Our approach will be to study poles and residues corresponding to the baryonic vevs of the monopole ancestors.

\subsubsection{The 4D RG flow and poles in the Index}
In this sub-section, we study the behavior of our formula for the Schur index under RG-flows triggered by turning on vevs for the 4D baryonic ancestors of the monopole operators discussed above. We expect that these 4D RG flows are in one-to-one correspondence with the flows discussed above in 3D and that the RG flows in 4D commute with the reduction along the circle.\footnote{This expectation is based on the fact that a non-perturbative superpotential is not compatible with eight supercharges.}

In what follows, we give a vacuum expectation value (vev) to a Higgs branch operator, $\mathcal{O}$. The authors of \cite{Gaiotto:2012xa} argued that the superconformal indices of the UV and IR SCFTs of such an RG flow are related to each other by
\begin{align}
\mathcal{I}_\text{vect}(q)^{-1}\cdot \mathcal{I}_\text{IR}(q;{\bf y}) = -f_{i,\mathcal{O}}\cdot\text{Res}_{x_i=x_i^*}\left(\frac{1}{x_i}\mathcal{I}_\text{UV}(q;{\bf x})\right)~,
\label{eq:index-relation}
\end{align}
where $f_{k,\mathcal{O}}$ is the flavor charge of $\mathcal{O}$ under the $k$-th flavor symmetry of the UV SCFT, and $x_i$ is a flavor fugacity associated with a flavor symmetry with non-vanishing $f_{i,\mathcal{O}}$. The contribution $\mathcal{I}_\text{vect}(q) \equiv [(q;q)_\infty]^2$ is the Schur index of a free vector multiplet. In \eqref{eq:index-relation}, $x_i^*$ is the value of $x_i$ such that
\begin{align}
q^{R_{\mathcal{O}}}\prod_{k=1}^{\text{rank}\, G_F}(x_k)^{f_{k,\mathcal{O}}} = 1~,
\end{align}
where $R_{\mathcal{O}}$ is the $SU(2)_R$ weight of $\mathcal{O}$, and $G_F$ is the flavor symmetry of the UV SCFT. We will see below that our formula for the Schur index of the $(A_{N-1},A_{N(n-1)-1})$ theory is perfectly consistent with the above index relation for various non-trivial RG flows.

\subsubsection{$(A_{N-1},A_{N(n-1)-1}) \to (A_{N-2},A_{(N-1)(n-1)-1})\oplus (A_1,A_1)$ flow}
Let us first study the RG-flow of the form $(A_{N-1},A_{N(n-1)-1}) \to (A_{N-2},A_{(N-1)(n-1)-1})\oplus (A_1,A_1)$. For generic $N$ and $n$ this is a flow between conformal manifolds in the UV and IR (with a decoupled axion-dilaton multiplet in the IR). To study this flow from the 4D index perspective, we first rewrite our formula for the index so that it is suitable for the above residue computation.

To that end, the expression for the $(A_{N-1},A_{N(n-1)-1})$ index that we proposed in \eqref{eq:general} can be rewritten as
\begin{align}
\mathcal{I}_{(A_{N-1},A_{N(n-1)-1})}(q;\bx) &= \frac{\prod_{k=1}^{N-1}(1-q^k)^{N-k}}{[(q;q)_\infty]^{2N-2}}\sum_{\lambda}q^{\frac{n}{2}(\lambda,\lambda+2\rho)}(\text{dim}_q\,R_\lambda)\,
 \text{Tr}_{R_\lambda}\left[q^{-\frac{n}{2}F^{ij}h_ih_j}\prod_{i=1}^{N-1}(x_1\cdots x_i)^{h_i}\right]~,
\label{eq:index1}
\end{align}
where $\lambda$ runs over the non-negative weights, and $\rho$ is the Weyl vector of $A_{N-1}$.\footnote{We use the fact that $C_2(R) = \frac{1}{2}(\lambda,\lambda + 2\rho)$.} Here $R_\lambda$ stands for the highest weight representation whose highest weight is given by $\lambda$. 

We first note that the quantum dimension of the representation $R_\lambda$ can be split into a $\lambda_{N-1}$-independent part and a $\lambda_{N-1}$-dependent one as follows\footnote{Here, $\ell_i$ is defined by $\ell_i = \sum_{j=i}^{N-1}\lambda_j$ as in footnote \ref{foot:ell}. }
\begin{align}
\text{dim}_q\,R_\lambda 
&= \text{dim}_q^{A_{N-2}} R_{(\lambda_1,\cdots,\lambda_{N-2})}^{A_{N-2}} \times \prod_{i=1}^{N-1}\frac{[\ell_i+N-i]_q}{[N-i]_q}~,
\label{eq:q-dim0}
\end{align}
where $R^{A_{N-2}}_{(\lambda_1,\cdots,\lambda_{N-2})}$ is the representation of $A_{N-2}$ corresponding to the dynkin label $(\lambda_1,\cdots,\lambda_{N-2})$ and the $q$-number is defined as
\begin{equation}\label{qnum}
\left[x\right]_q\equiv{q^{-{x\over2}}-q^{x\over2}\over q^{-{1\over2}}-q^{1\over2}}~.
\end{equation}
Now, note that the last factor on the RHS of \eqref{eq:q-dim0} can be expanded as follows
\begin{align}
\prod_{i=1}^{N-1}\frac{[\ell_i+N-i]_q}{[N-i]_q} = \frac{(-1)^{N-1}q^{-\frac{N-1}{2}\lambda_{N-1}-\frac{1}{2}\sum_{k=1}^{N-2}k\lambda_k-\frac{N(N-1)}{4}} + (\text{higher powers of }q^{\lambda_{N-1}} )}{\prod_{i=1}[N-i]_q}~.
\label{eq:q-dim}
\end{align}
Furthermore, since $F^{ij}h_ih_j$ is Weyl invariant, we have\footnote{As usual, $|\eta|^2 \equiv (\eta,\eta)$ and $\vec{n}\cdot \vec{\alpha} \equiv \sum_{i=1}^{N-1}n_i\alpha_i$.}
\begin{align}
\text{Tr}_{R_\lambda}\left[q^{-\frac{n}{2}F^{ij}h_ih_j}\prod_{i=1}^{N-1}(x_1\cdots x_i)^{h_i}\right] &= \sum_{\vec{n}\in M_\lambda}q^{-\frac{n}{2}|\lambda-\vec{n}\cdot\vec{\alpha}|^2}\sum_{\mu\in W_{\lambda-\vec{n}\cdot\vec{\alpha}}}\prod_{i=1}^{N-1}(x_1\cdots x_i)^{\mu_i}~,
\label{eq:trace1}
\end{align}
where $M_\lambda \subset\mathbb{N}^{N-1}$ is the set of non-negative integers, $\vec{n}$, such that $(\lambda-\vec{n}\cdot \vec{\alpha})_i$ is non-negative for all $i$, and $W_{\lambda-\vec{n}\cdot\vec{\alpha}}$ is the Weyl orbit including the weight $\lambda-\vec{n}\cdot\vec{\alpha}$.\footnote{Here we used the fact that the multiplicity of $\lambda-\vec{n}\cdot\vec{\alpha}$ in the representation $R_\lambda$ is the number of ways of writing $\lambda-\vec{n}\cdot\vec{\alpha}$ as a linear combination of positive roots. If $(\lambda-\vec{n}\cdot\vec{\alpha})_i\neq 0$ for all $i$, then the Weyl orbit, $W_{\lambda-\vec{n}\cdot\vec{\alpha}}$, contains $N!$ weights, while, if $(\lambda-\vec{n}\cdot\vec{\alpha})_i=0$ for some $i$, then there are fewer weights.}  
Combining the expressions \eqref{eq:q-dim0}--\eqref{eq:trace1}, we obtain
\begin{align}
&\mathcal{I}_{(A_{N-1},A_{N(n-1)-1})}(q;{\bf x}) = \frac{\prod_{k=1}^{N-1}(1-q^k)^{N-k}}{[(q;q)_\infty]^{2N-2}}\frac{(-1)^{N-1}}{\prod_{i=1}^{N-1}[N-i]_q}\sum_{\lambda}\sum_{\vec{n}\in M_\lambda} \text{dim}_q^{A_{N-2}}R_{(\lambda_1,\cdots,\lambda_{N-2})}^{A_{N-2}}
\nonumber\\
& \times \left(q^{\left(\frac{(n-1)(N-1)}{2}+nn_{N-1}\right)\lambda_{N-1}-\frac{1}{2}\sum_{k=1}^{N-2}k\lambda_k-\frac{N(N-1)}{4}} + (\text{higher powers of }q^{\lambda_{N-1}} )\right)
\nonumber\\
&\times q^{\frac{n}{2}\sum_{i=1}^{N-2}i(N-i)\lambda_i+n\sum_{i=1}^{N-2}\lambda_i n_i- \frac{n}{2}\left|\vec{n}\cdot\vec{\alpha}\right|^2}\left( \prod_{k=1}^{N-1}(x_1\cdots x_{k})^{\lambda_k +n_{k-1}-2n_{k}+n_{k+1}}  + \text{Weyl conjugates}\right)~,
\label{eq:residue1}
\end{align}
where we defined $n_{0},\,n_{N}\equiv 0$.
Note here that
\begin{align} 
\sum_{\lambda}\sum_{\vec{n}\in M_\lambda} &= \sum_{n_1,\cdots,n_{N-1}=0}^\infty \sum_{\lambda \in \tilde{M}_{\vec{n}}}~,
\end{align}
where $\tilde{M}_{\vec{n}}$ is the set of weights $\lambda$ such that $(\lambda - \vec{n}\cdot \vec{\alpha})_i\geq 0$ for all $i=1,\cdots,N-1$. For a given $\vec{n}$, while there is no maximum value of $\lambda_{N-1}$ so that $\lambda\in \tilde{M}_{\vec{n}}$, there exists a minimum value, say $\lambda_{N-1}^*$. Then we see that \eqref{eq:residue1} contains the following sum
\begin{align}
&\sum_{\lambda_{N-1}=\lambda_{N-1}^*}^\infty \bigg(q^{\left(\frac{(n-1)(N-1)}{2}+nn_{N-1}\right)\lambda_{N-1} -\frac{1}{2} \sum_{k=1}^{N-2}k\lambda_k -\frac{N(N-1)}{4}} + (\text{higher powers of }q^{\lambda_{N-1}} )\bigg)
\nonumber\\
& \qquad \times  \left(\prod_{k=1}^{N-1}(x_1\cdots x_{k})^{\lambda_k +n_{k-1}-2n_{k}+n_{k+1}} + \text{Weyl conjugates}\right)~.
\end{align}
Since the sum over $\lambda_{N-1}$ gives a geometric series in $x_1x_2\cdots x_{N-1}q^{\frac{(N-1)(n-1)}{2}+nn_{N-1}}$, the index has a pole at $x_1x_2\cdots x_{N-1}q^{\frac{(N-1)(n-1)}{2}+nn_{N-1}}=1$. In particular, the index has a pole at
\begin{align}
x_1x_2\cdots x_{N-1}q^{\frac{(N-1)(n-1)}{2}}=1~,
\label{eq:pole}
\end{align} 
which arises from the contributions with $n_{N-1}=0$. The corresponding residue is evaluated as
\begin{align}
-\text{Res}\;\mathcal{I}_{(A_{N-1},A_{N(n-1)-1})}(q;{\bf x})
&= \left[\mathcal{I}_\text{vect}(q)\right]^{-1}\cdot  \mathcal{I}_{(A_{N-2},A_{(N-1)(n-1)-1})}(q;{\bf y})~,
\label{eq:res1}
\end{align}
where $y_i\equiv x_iq^{\frac{n-1}{2}}$.\footnote{Here we used the fact that $\lambda_{N-1}^*=0$ for $n_{N-1}=0$.} Note that the condition \eqref{eq:pole} now reduces to
\begin{align}
y_1\cdots y_{N-1}=1~,
\end{align}
which is required for the Weyl symmetry of $A_{N-2}$ that corresponds to the permutations of $y_1,\cdots, y_{N-1}$.
The result \eqref{eq:res1} agrees with the expected index relation \eqref{eq:index-relation} for the RG-flow
\begin{equation}
(A_{N-1},A_{N(n-1)-1}) \to (A_{N-2},A_{(N-1)(n-1)-1})\oplus (A_1,A_1)~,
\end{equation}
and is compatible with the discussion in Sec.~\ref{3DRG} from the perspective of the dimensionally reduced mirror. This is a highly non-trivial check of our formula for the Schur index of the $(A_{N-1},A_{N(n-1)-1})$ theory and also of the proposed mirror in \cite{Xie:2012hs,Xie:2013jc}.

\subsubsection{The $(A_{N-1},A_{N(n-1)-1}) \to (A_{\nu-1},A_{\nu(N-1)-1})\oplus(A_{N-\nu-1},A_{(N-\nu)(n-1)-1})\oplus (A_1,A_1)$ flow}
It is straightforward to generalize the above discussion to the more general RG flow
\begin{equation}
(A_{N-1},A_{N(n-1)-1}) \to (A_{\nu-1},A_{\nu(N-1)-1})\oplus(A_{N-\nu-1},A_{(N-\nu)(n-1)-1})\oplus (A_1,A_1)~, \ \ \ 1<\nu<N-1~.
\end{equation}
For generic values of $N$ and $n$, this is an RG flow from a UV conformal manifold to an IR theory that is a direct product of two smaller conformal manifolds and a decoupled theory (comprising the axion-dilaton multiplet). The 3D descendant of this RG flow was described above.

To reproduce this flow from the index, we first note that the quantum dimension \eqref{eq:q-dim} can also be rewritten as\footnote{Recall here that $\ell_i \equiv \sum_{k=i}^{N-1}\lambda_k$.}
\begin{align}
 \text{dim}_q\,R_\lambda &= \text{dim}_q\,R_{(\lambda_1,\cdots,\lambda_{\nu-1})}^{A_{\nu-1}}\times \text{dim}_q\,R_{(\lambda_{\nu+1},\cdots,\lambda_{N-1})}^{A_{N-\nu-1}} \times \prod_{1\leq i\leq \nu<j\leq N} \frac{[\ell_i-\ell_j+j-i]_q}{[j-i]_q}~.
\end{align}
Moreover, the last factor on the RHS can be expanded as follows
\begin{align}
&
(-1)^{\nu(N-\nu)}\frac{q^{-\frac{1}{2}\left(\nu(N-\nu)\lambda_\nu + (N-\nu)\sum_{k=1}^{\nu-1}k{\lambda_k} +\nu\sum_{k=\nu+1}^{N-1}(N-k)\lambda_{k} + \frac{1}{2}N\nu(N-\nu)\right)}}{\prod_{1\leq i\leq \nu<j\leq N}[j-i]_q} +\; \left(\text{higher powers of }q^{\lambda_{\nu}}\right)~.
\end{align}
Combining these expressions with \eqref{eq:trace1}, 
we will see that the sum over $\lambda_\nu$ gives a geometric series in $x_1\cdots x_{\nu}q^{\frac{(n-1)\nu(N-\nu)}{2}+n n_\nu}$ and therefore the index $\mathcal{I}_{(A_{N-1},A_{N(n-1)-1})}(q;{\bf x})$ has a pole at $x_1\cdots x_\nu q^{\frac{(n-1)\nu(N-\nu)}{2}+n n_\nu}=1$ for all $n_\nu = 0,1,2,3,\cdots$. In particular, the index has a pole at
\begin{align}
 x_1\cdots x_{\nu}\, q^{\frac{(n-1)\nu(N-\nu)}{2}}=1~,
\label{eq:cond}
\end{align}
with the residue given by
\begin{align}
-\text{Res}\;\mathcal{I}_{(A_{N-1},A_{N(n-1)-1})}(q;{\bf x}) 
&= [\mathcal{I}_\text{vect}(q)]^{-1}\cdot\mathcal{I}_{(A_{\nu-1},A_{\nu(n-1)-1})}(q;{\bf y})\cdot \mathcal{I}_{(A_{N-\nu-1},A_{(N-\nu)(n-1)-1})}(q;{\bf z})~.
\label{eq:RG2}
\end{align}
Here, the flavor fugacities on the RHS are defined as $y_i \equiv x_i q^{\frac{(n-1)(N-\nu)}{2}}$ for $i=1,\cdots,\nu$ and $z_i\equiv x_{\nu+i} q^{-\frac{(n-1)\nu}{2}}$ for $i=1,\cdots,N-\nu$. Note that \eqref{eq:cond} and $x_1\cdots x_{N}=1$ imply
\begin{align}
\prod_{i=1}^{\nu} y_i = 1~,\qquad \prod_{i=1}^{N-\nu}z_i = 1~.
\end{align}
The relation \eqref{eq:RG2} agrees with the expected index relation \eqref{eq:index-relation} for the RG-flow, $(A_{N-1},A_{N(n-1)-1}) \to (A_{\nu-1},A_{\nu(n-1)-1})\oplus (A_{N-\nu-1},A_{(N-\nu)(n-1)-1})\oplus (A_1,A_1)$. This result is also compatible with the 3D discussion in Sec.~\ref{3DRG}.

\subsection{Cardy-like behavior}
In this section, we would like to study the essential singularity that arises in the index when we take $q\to1$. On general grounds, we expect that the Schur index behaves as follows in this limit \cite{Buican:2015ina,Ardehali:2015bla,DiPietro:2016ond,DiPietro:2014bca}
\begin{equation}\label{Schurasy}
\lim_{\beta\to0}\log\mathcal{I}_{\rm Schur}=-{8\pi^2\over\beta}(a-c)+\cdots~,
\end{equation}
where $\beta$ is defined as $q=e^{-\beta}$ and is proportional to the $S^1$ radius (when thinking of the index as being related to the twisted partition function of the theory on $S^3\times S^1$).\footnote{A more rigorous statement is that if the $S^3$ partition function of the theory reduced on $S^1$ is finite, then \eqref{Schurasy} holds \cite{Ardehali:2015bla,DiPietro:2016ond} (although we are not aware of any $\CN=2$ SCFT counterexamples to the behavior in \eqref{Schurasy}). The $(A_{N-1}, A_{N(n-1)-1})$ theories we are considering in this note satisfy the above criteria: their mirror duals consist of $N-1$ $U(1)$ gauge groups with $n-1$ fundamentals for each $U(1)$ and $n-1$ bifundamentals between each pair of abelian nodes.}

Furthermore, if the theory in question has a genuine Higgs branch (i.e., a branch on which there are, at generic points, just free hypermultiplets), then $U(1)_R$ 't Hooft anomaly matching guarantees that
\begin{equation}\label{higgs}
\lim_{\beta\to0}\log\mathcal{I}_{\rm Schur}={\pi^2\over3\beta}\dim_{\mathbb{Q}}\mathcal{M}_{H}~,
\end{equation}
where $\dim_{\mathbb{Q}}\mathcal{M}_H$ is the quaternionic dimension of the Higgs branch, $\CM_H$.

Now, our conjecture in \eqref{eq:general} is a sum over products of terms whose power is fixed by the topology of $\CC$ (which we will refer to as the \lq\lq topological" terms) \eqref{topological} and components of irregular wavefunctions \eqref{eq:wf1}. It is clear that the only terms that contribute to the pole in \eqref{Schurasy} are those involving factors of $(q;q)^{-1}$. The topological terms supply $N-1$ such factors, and the irregular wavefunction supplies an additional $N-1$ factors. Each such factor is just the $q\to1$ contribution of a half-hypermultiplet. Therefore, we have
\begin{equation}\label{cardyI}
\lim_{\beta\to0}\log\mathcal{I}_{(A_{N-1},A_{N(n-1)-1})}={\pi^2\over3\beta}(N-1)+\cdots~,
\end{equation}
which reproduces the known $a-c$ for these theories given by taking the difference of the expressions in \eqref{acAD}.

As a simple application to see how easily our results extend to the case of type IV AD theories \cite{Xie:2013jc} (recall from the introduction that these theories are characterized by an additional regular singularity), let us turn our attention to the $\mathcal{I}_{(I_{N,N(n-1)}, F)}$ SCFT (here \lq\lq $F$" refers to the fact that the regular puncture is \lq\lq full"). The irregular wave function is the same as in the $(A_{N-1}, A_{N(n-1)-1})$ case and therefore again contributes a factor of ${\pi^2(N-1)\over6\beta}$ to \eqref{Schurasy}. On the other hand, the full regular puncture corresponds to the following Young diagram: $[1,\cdots ,1]$. The resulting flavor symmetry is $SU(N)$. The adjoint representation has dimension $N^2-1$, and so we find that
\begin{equation}\label{cardyI3}
\lim_{\beta\to0}\log\mathcal{I}_{(I_{N,N(n-1)},F)}={\pi^2\over6\beta}(N^2+N-2)+\cdots={\pi^2\over6\beta}(N+2)(N-1)+\cdots~,
\end{equation}
which matches the known results in (2.43) of \cite{Xie:2013jc}.

\section{Discrete Symmetries of the Index and $S$-duality}

As we discussed in Sec. \ref{subsec:motivation}, our wave function
\begin{align}
\tilde{f}_{R}^{(n)}(q;{\bf x}) &= \prod_{k=1}^\infty\left(\frac{1}{1-q^k}\right)^{N-1}q^{nC_2(R)}\text{Tr}_R\left[q^{-\frac{n}{2}F^{ij}h_ih_j}\prod_{i=1}^{N-1}(x_1\cdots x_i)^{h_i}\right]~,
\end{align}
for the irregular singularity is invariant under the action of Weyl group of $SU(N)$ on the flavor fugacities, $x_i$, even though the flavor symmetry associated with the irregular puncture is not $SU(N)$ but $U(1)^{N-1}$. More concretely, the above wave function is invariant under
\begin{align}
s_i: x_i\leftrightarrow x_{i+1}
\label{eq:Weyl}
\end{align}
for $i=1,\cdots, N-1$, which generate $S_N$ acting on the flavor fugacities, ${\bf x}$.\footnote{Recall here that $x_{N}\equiv (x_1\cdots x_{N-1})^{-1}$. With this definition, $s_i$ corresponds to the Weyl reflection $\lambda_k \to \lambda_k - (\alpha_i,\alpha_k)\lambda_i$. } This invariance implies that the Schur indices of the $(A_{N-1},A_{N(n-1)-1})$ and $(I_n, R_{Y})$ Arygres-Douglas theories are invariant under the $S_{N}$ action on the flavor fugacities.

This $S_N$ symmetry of the index is consistent with the fact that our irregular puncture is type I and therefore corresponds to the boundary condition \eqref{irregsing} of the Hitchin system. To see this, let us collect the singular terms in \eqref{irregsing} as
\begin{align}
M(z) \equiv M_1z^{n-1} + M_2z^{n-2} + \cdots M_n + \frac{M_{n+1}}{z}~.
\label{eq:boundary}
\end{align}
Since we can diagonalize this matrix order by order without changing the spectral curve, $\det (xdz - \varphi(z))=0$, we assume the $M_i$ are all diagonal. Then the (diagonal) elements of the $M_i$ are the coupling constants and mass parameters of the correspoinding 4d $\mathcal{N}=2$ theory. In particular, the parameters in $M_1$ correspond to exactly marginal couplings and those in $M_{n+1}$ are mass parameters, while those in $M_i$ for $2\leq i\leq n$ correspond to relevant couplings. Note here that the only constraint on $M_i$ is that the sum of its eigenvalues is vanishing. Therefore there is an action of $S_N$ that permutes the $N$ eigenvalues of $M(z)$. Since such an $S_N$ action can be realized by a gauge transformation in the Hitchin system,\footnote{We can make this gauge transformation consistent with a possible regular puncture at $z=0$.} it preserves the Seiberg-Witten curve, $\det(xdz -\varphi(z))=0$, of the 4d $\mathcal{N}=2$ theory up to re-labeling the Coulomb branch parameters.\footnote{The Coulomb branch parameters of the 4d theory correspond to the Hitchin moduli that are not fixed by the boundary condition at punctures. Since they correspond to the vacuum expectation values of Coulomb branch operators, this re-labeling of the Coulomb branch parameters corresponds to a change of a basis of the Coulomb branch chiral ring.} In the same spirit as \cite{Seiberg:1994aj, Gaiotto:2009we, Buican:2014hfa}, this discussion suggests that the $(A_{N-1},A_{N(n-1)-1})$ and $(I_{N,N(n-1)},R_Y)$ Argyres-Douglas theories are invariant under this $S_N$.
Note here that, since it permutes the diagonal elements of $M_1$, this $S_N$ symmetry generically relates the theory at different points on the conformal manifold. Therefore this $S_N$ contains an S-duality group of the theory as a sub-group.\footnote{In the case of an additional regular singularity, we may also have additional discrete symmetry (and part of this discrete symmetry may also be part of the $S$-duality group).}

This $S_N$ action on the boundary condition \eqref{eq:boundary} can naturally be identified with the $S_N$ action \eqref{eq:Weyl} on the flavor fugacities, since a permutation of the mass parameters encoded in $M_{n+1}$ corresponds to a change of the basis of flavor charges. Then, the $S_N$ invariance of the index is consistent with the $S_N$ symmetry of the Seiberg-Witten curve. Indeed, in Sec.~\ref{subsec:lower}, we have seen that the $S_4$ invariance of the $(A_3,A_3)$ index and the $S_3$ invariance of the $(A_2,A_5)$ index are related to the S-duality discussed in \cite{Buican:2014hfa, Buican:2015tda}. This identification of the two $S_N$ will tell us how the Schur operators of general $(A_{N-1},A_{N(n-1)-1})$ and $(I_{N,N},R_{Y})$ theories are mapped by the S-duality. This will be useful to identify the 2d chiral algebras associated, in the sense of \cite{Beem:2013sza}, with these Argyres-Douglas theories.

As a final point, we note that the index is also invariant under a $\mathbb{Z}_2$ charge conjugation
\begin{align}
x_1\to (x_1)^{-1}~, \quad x_2 \to (x_2)^{-1}~,\quad x_3\to (x_3)^{-1}~,\quad \cdots~,\quad x_{N-1}\to (x_{N-1})^{-1}~.
\end{align}
Combined with the above $S_N$ symmetry, we see that, for generic $N$, the discrete symmetry of the index includes $S_N\times\mathbb{Z}_2$ (for $N=2$, this group is reduced to $\mathbb{Z}_2$). However, the discrete symmetry group acting on the index may be larger (as it will generally be when there is also a regular puncture). To study such a possibility, it is useful to recall \eqref{flavor}. A more general discrete symmetry acting only on ${\bf x}$ and preserving $C_R(q)$ can  interchange different components of the wavefunctions, $\tilde f^{(n)}_{R_{\lambda_i}}(q;{\bf x})$ ($i=1,\cdots,N'$), only if $(\lambda_1,\rho)=(\lambda_2,\rho)=\cdots=(\lambda_{N'},\rho)$. In the case of conjugate representations, this latter quantity is the same. However, it would be interesting to see if there are more general group actions that are consistent with this constraint and are indeed symmetries of the index.

\section{Beyond $A_{N-1}$?}\label{speculation}
One amusing aspect of our wave function formula in \eqref{eq:wf1} is that it can, in principle, be defined for any Lie algebra, $\mathfrak{g}$, by a mild re-writing
\begin{equation}\label{moregen}
\tilde{f}_{R}^{(n)}(q;{\bf x}) =\prod_{k=1}^\infty\left(\frac{1}{1-q^k}\right)^{r_{\mathfrak{g}}}q^{nC_2(R)}\text{Tr}_R\left[q^{-\frac{n}{2}F^{ij}h_ih_j}\prod_{i=1}^{N-1}(x_1\cdots x_i)^{h_i}\right]~,
\end{equation}
where $r_{\mathfrak{g}}$ is the rank of $\mathfrak{g}$, $F^{ij}$ is the quadratic form for $\mathfrak{g}$, and the $h_i$ are the Cartans for $\mathfrak{g}$.\footnote{Of course, the above generalization is not unique. For example, we could take $r_{\mathfrak{g}}\to h_{\mathfrak{g}}^{\vee}-1$ in \eqref{moregen} and below (where $h_{\mathfrak{g}}^{\vee}$ is the dual Coxeter number). Note that for generic $\mathfrak{g}$, $h^{\vee}_{\mathfrak{g}}-1\ne r_{\mathfrak{g}}$. The choice we make is somewhat more natural: it guarantees that the rank of the flavor symmetry (as defined by the number of flavor-singlet terms in the index at $\CO(q)$) is the same as the number of fugacities in \eqref{moregen}.} We can also try to generalize \eqref{topological} as follows
\begin{equation}\label{rewriteCr}
C_R(q)=\prod_{k=1}^{r_{\mathfrak{g}}}\prod_{i=1}^{\infty}{1\over1-q^{d_k-1+i}}{\rm dim}_qR~,
\end{equation}
where, in the regular puncture case with simply laced $\mathfrak{g}$, the authors of \cite{Mekareeya:2012tn,Lemos:2012ph} argued that the ${\bf d}_{\mathfrak{g}}=\left\{d_1,\cdots,d_{r_{\mathfrak{g}}}\right\}$ are the degrees of invariants of $\mathfrak{g}$,\footnote{Since this is a topological factor, it is likely to apply to the irregular puncture case as well.} and ${\rm dim}_qR$ is defined as
\begin{equation}\label{genqdim}
{\rm dim}_qR\equiv\prod_{\alpha>0}{\left[(\lambda+\rho,\tilde\alpha)\right]_q\over\left[(\rho,\tilde\alpha)\right]_q}~, \ \ \ \tilde\alpha\equiv{2\alpha\over(\theta,\theta)}~.
\end{equation}
Here $\lambda$ are the Dynkin labels, $\alpha$ is any positive root, $\rho$ is the Weyl vector, $\theta$ is the highest root, and $[x]_q$ was defined in \eqref{qnum}.

Given these expressions, we can attempt to construct a putative Schur index
\begin{equation}\label{genind}
\mathcal{I}_{(\mathfrak{g},n)}(q;{\bf x}) = \sum_{R}C_R(q)\tilde{f}^{(n)}_R(q;{\bf x})~,
\end{equation}
where, for $\mathfrak{g}=A_{N-1}$, the hypothetical $(\mathfrak{g},n)$ SCFT is the actual $(A_{N-1},A_{N(n-1)-1})$ theory. If such expressions are meaningful for more general $\mathfrak{g}$, they may not come from a $(2,0)$ SCFT compactified on $\CC$.\footnote{If $\mathfrak{g}\ne A,D,E$, there are various obstructions to a $(2,0)$ origin of the theory (e.g., see \cite{Cordova:2015vwa} for an interesting recent discussion). Moreover, even if $\mathfrak{g}=D,E$, we are not sure which SCFTs---if any---\eqref{moregen}, \eqref{rewriteCr}, and \eqref{genind} correspond to.}  Therefore, it is not completely clear what the rules are (or even if they exist!) for writing an index given the data in \eqref{moregen} and \eqref{rewriteCr}. However, we can simply press on and ask if \eqref{genind} satisfies the basic rules for a Schur index.

The most obvious generalizations to check are those involving $\mathfrak{g}=D_N, E_{6,7,8}$ (where there might be some hope that, at least for some paris $(\mathfrak{g},n)$, a theory of class $\CS$ can be associated with \eqref{moregen}---see the discussion in the regular puncture case given in \cite{Mekareeya:2012tn,Lemos:2012ph}). We defer such a discussion to future work and instead focus on an even more speculative avenue with non-simply laced $\mathfrak{g}$.

In this spirit, one can check that for $\mathfrak{g}=G_2$ and $n=2$, the expression in \eqref{genqdim} cannot correspond to a Schur index. Indeed, plugging the following group theory data
\begin{equation}
F_{G_2}=\left[\begin{array}{cc}
{2\over3} & 1 \\
1 & 2\\
\end{array}\right]~, \ \ \ {\bf d}_{G_2}=\left\{2, 6\right\}~,
\end{equation}
into \eqref{moregen} results in powers of $q$ that are not integer or half-integer. Clearly, we must, at the very least, specialize to a subset of $\mathfrak{g}$ and $n$.

With these facts in mind, consider for example the case $\mathfrak{g}=F_4$ and $n=2$ with
\begin{equation}
F_{F_4}=\left[\begin{array}{cccc}
2 & 3 & 2 & 1 \\
3 & 6 & 4 & 2 \\
2 & 4 & 3 & {3\over2} \\
1 & 2 & {3\over2} & 1\\
\end{array}\right]~, \ \ \ {\bf d}_{F_4}=\left\{2, 6, 8, 12\right\}~.
\end{equation}
It is straightforward to verify that all powers appearing in \eqref{genind} are integer or half-integer and that the expansion in $q$ of this index reads
\begin{equation}\label{F4index}
\CI_{(F_4,2)}(q;{\bf x})=1+4q+15q^2+45q^3+125q^4+316q^5+\CO(q^{11\over2})~,
\end{equation}
where the first flavor dependence comes in at $\CO(q^{11\over2})$. If the corresponding theory exists, it has flavor symmetry $U(1)^4$ and has a candidate stress tensor contribution at $\CO(q^2)$. Note that this theory cannot be the putative $F_4$ SCFT discussed in \cite{Beem:2013sza,Shimizu:2017kzs} since the flavor symmetry here is not $F_4$ but rather is its Cartan subalgebra (note however, that the order 1152 Weyl group of $F_4$ must still be a symmetry of \eqref{F4index}).\footnote{Moreover, one would guess from our formula that the hypothetical Higgs branch has quaternionic dimension 4 instead of 8.} It would be interesting to see if one can bootstrap a chiral algebra (with a finite number of generators) whose vacuum character satisfies \eqref{F4index} and gives the Schur sector of a genuine 4D $\CN=2$ SCFT.\footnote{Note that the chiral algebra for the hypothetical $(F_4,2)$ SCFT would need to include new AKM primaries at $\CO(q^{11\over2})$.}

\section{Discussion and Conclusion}
From our simple proposal for the wave function for an irregular singularity of type I in $SU(N)$ $q$-deformed Yang-Mills theory \eqref{eq:wf1}, we extracted a great deal of physics concerning new superconformal indices, conformal manifolds, $S$-duality, RG flows, and $S^1$ reductions. Based on this discussion, we suggest the following open problems:

\begin{itemize}
\item It would be interesting to find the chiral algebras associated with the $(A_{N-1}, A_{N(n-1)-1})$ and $(I_{N,N(n-1)}, R_Y)$ SCFTs (see \cite{Creutzig:2017qyf,BR} for a beautiful recent discussion of the $(A_1, A_{2n-3})$ and $(I_{2,2(n-1)}, F_{A_1})$ cases). Our discussion implies there is a non-trivial $S_N\times\mathbb{Z}_2$ action on the associated chiral algebra.
\item It would be worthwhile to understand the full symmetry group that acts on the index and which subgroup corresponds to the action of the $S$-duality group. In so doing, perhaps we can make closer contact with the results in \cite{Caorsi:2016ebt}.
\item Since we can compute the $(A_2, A_2)\sim (A_1, D_4)$ index either as a correlator in $SU(3)$ $q$-deformed Yang-Mills on a surface with a single irregular puncture or as a correlator in $SU(2)$ $q$-deformed Yang-Mills theory with both a regular and an irregular puncture, we see that certain observables in different $q$-deformed Yang-Mills theories (on different Riemann surfaces) must be related. Can this result be promoted to a duality between different subsectors of these theories?
\item Can our expressions for wavefunctions in the $A_{N-1}$ case be extended to other Lie algebras and give information about new SCFTs (perhaps even making contact with some of the results in \cite{Argyres:2016xua} or their generalizations to higher ranks)? We made a speculative proposal in this regard and briefly discussed the hypothetical case of an SCFT associated with $F_4$ (and $n=2$) (note again that this theory is not the hypothetical $F_4$ SCFT discussed in \cite{Beem:2013sza,Shimizu:2017kzs} since it only has $U(1)^4\subset F_4$ flavor symmetry, although it does have the full discrete Weyl symmetry, $W(F_4)$). The $D_N$ and $E_{6,7,8}$ cases may be even more promising avenues for study. Can one find an associated chiral algebra and 4D $\CN=2$ theory using certain bootstrap techniques? Even though our expressions typically don't manifest the full $\mathfrak{g}$ flavor symmetry, they are still invariant under the (sometimes very large) Weyl groups, $W(\mathfrak{g})$, which may furnish strong constraints on the resulting algebras.
\item We described some intricate RG flows between conformal manifolds in the UV and IR. It would be interesting to understand if these flows obey any new constraints beyond the $a$-theorem.
\end{itemize}

\ack{ \bigskip
M.~B. would like to thank the ICTP and SISSA theory groups for stimulating discussions and hospitality during the completion of this work. M.~B. would also like to thank the Aspen Center for Physics (NSF grant \#PHY-1066293) and Perimeter Institute (supported by the Government of Canada and the Province of Ontario) for hosting interesting workshops during the course of this research. M.~B.'s research is partially supported by the Royal Society under the grant \lq\lq New Constraints and Phenomena in Quantum Field Theory." T.~N.'s research is partially supported by JSPS Grant-in-Aid for Scientific Research (B) No. 16H03979. 
}

\newpage
\bibliography{chetdocbib}

\begin{thebibliography}{10}
\ifx\href\asklfhas\newcommand{\href}[2]{#2}\fi
\ifx\arxivref\asklfhas\newcommand{\arxivref}[2]{\href{http://arxiv.org/abs/#1}{#2}}\fi
\ifx\doiref\asklfhas\newcommand{\doiref}[2]{\href{http://dx.doi.org/#1}{#2}}\fi
\parskip 0pt
\normalsize

\bibitem{Gadde:2011ik}
A.~Gadde, L.~Rastelli, S.~S. Razamat \& W.~Yan,
\textit{``{The 4d Superconformal Index from q-deformed 2d Yang-Mills}''},
\doiref{10.1103/PhysRevLett.106.241602}{Phys.~Rev.~Lett. \textbf{106}, 241602
  (2011)\ignorespaces}\ignorespaces,
\normalsize{\texttt{\arxivref{1104.3850}{arXiv:1104.3850}}}\ignorespaces
\bibitem{Buican:2015ina}
M.~Buican \& T.~Nishinaka,
\textit{``{On the superconformal index of Argyres–Douglas theories}''},
\doiref{10.1088/1751-8113/49/1/015401}{J.~Phys. \textbf{A49}, 015401
  (2016)\ignorespaces}\ignorespaces,
\normalsize{\texttt{\arxivref{1505.05884}{arXiv:1505.05884}}}\ignorespaces
\bibitem{Cecotti:2010fi}
S.~Cecotti, A.~Neitzke \& C.~Vafa,
\textit{``{R-Twisting and 4d/2d Correspondences}''},
\normalsize{\texttt{\arxivref{1006.3435}{arXiv:1006.3435}}}\ignorespaces
\bibitem{Xie:2012hs}
D.~Xie,
\textit{``{General Argyres-Douglas Theory}''},
\doiref{10.1007/JHEP01(2013)100}{JHEP \textbf{1301}, 100
  (2013)\ignorespaces}\ignorespaces,
\normalsize{\texttt{\arxivref{1204.2270}{arXiv:1204.2270}}}\ignorespaces
\bibitem{Argyres:1995jj}
P.~C. Argyres \& M.~R. Douglas,
\textit{``{New phenomena in SU(3) supersymmetric gauge theory}''},
\doiref{10.1016/0550-3213(95)00281-V}{Nucl.~Phys. \textbf{B448}, 93
  (1995)\ignorespaces}\ignorespaces,
\normalsize{\texttt{\arxivref{hep-th/9505062}{hep-th/9505062}}}\ignorespaces
\bibitem{Argyres:1995xn}
P.~C. Argyres, M.~R. Plesser, N.~Seiberg \& E.~Witten,
\textit{``{New N=2 superconformal field theories in four-dimensions}''},
\doiref{10.1016/0550-3213(95)00671-0}{Nucl.~Phys. \textbf{B461}, 71
  (1996)\ignorespaces}\ignorespaces,
\normalsize{\texttt{\arxivref{hep-th/9511154}{hep-th/9511154}}}\ignorespaces
\bibitem{Eguchi:1996vu}
T.~Eguchi, K.~Hori, K.~Ito \& S.-K. Yang,
\textit{``{Study of N=2 superconformal field theories in four-dimensions}''},
\doiref{10.1016/0550-3213(96)00188-5}{Nucl.~Phys. \textbf{B471}, 430
  (1996)\ignorespaces}\ignorespaces,
\normalsize{\texttt{\arxivref{hep-th/9603002}{hep-th/9603002}}}\ignorespaces
\bibitem{Gadde:2011uv}
A.~Gadde, L.~Rastelli, S.~S. Razamat \& W.~Yan,
\textit{``{Gauge Theories and Macdonald Polynomials}''},
\doiref{10.1007/s00220-012-1607-8}{Commun.Math.Phys. \textbf{319}, 147
  (2013)\ignorespaces}\ignorespaces,
\normalsize{\texttt{\arxivref{1110.3740}{arXiv:1110.3740}}}\ignorespaces
\bibitem{Beem:2013sza}
C.~Beem, M.~Lemos, P.~Liendo, W.~Peelaers, L.~Rastelli \& B.~C. van~Rees,
\textit{``{Infinite Chiral Symmetry in Four Dimensions}''},
\doiref{10.1007/s00220-014-2272-x}{Commun.~Math.~Phys. \textbf{336}, 1359
  (2015)\ignorespaces}\ignorespaces,
\normalsize{\texttt{\arxivref{1312.5344}{arXiv:1312.5344}}}\ignorespaces
\bibitem{Cordes:1994fc}
S.~Cordes, G.~W. Moore \& S.~Ramgoolam,
\textit{``{Lectures on 2-d Yang-Mills theory, equivariant cohomology and
  topological field theories}''},
\doiref{10.1016/0920-5632(95)00434-B}{Nucl.~Phys.~Proc.~Suppl. \textbf{41}, 184
  (1995)\ignorespaces}\ignorespaces,
\normalsize{\texttt{\arxivref{hep-th/9411210}{hep-th/9411210}}}\ignorespaces,
in \textit{``{NATO Advanced Study Institute: Les Houches Summer School, Session
  62: Fluctuating Geometries in Statistical Mechanics and Field Theory Les
  Houches, France, August 2-September 9, 1994}''},
184-244\ignorespaces
\bibitem{deHaro:2006uvl}
S.~de~Haro, S.~Ramgoolam \& A.~Torrielli,
\textit{``{Large N expansion of q-deformed two-dimensional Yang-Mills theory
  and Hecke algebras}''},
\doiref{10.1007/s00220-007-0232-4}{Commun.~Math.~Phys. \textbf{273}, 317
  (2007)\ignorespaces}\ignorespaces,
\normalsize{\texttt{\arxivref{hep-th/0603056}{hep-th/0603056}}}\ignorespaces
\bibitem{Kimura:2008gs}
Y.~Kimura \& S.~Ramgoolam,
\textit{``{Holomorphic maps and the complete 1/N expansion of 2D SU(N)
  Yang-Mills}''},
\doiref{10.1088/1126-6708/2008/06/015}{JHEP \textbf{0806}, 015
  (2008)\ignorespaces}\ignorespaces,
\normalsize{\texttt{\arxivref{0802.3662}{arXiv:0802.3662}}}\ignorespaces
\bibitem{Szabo:2013vva}
R.~J. Szabo \& M.~Tierz,
\textit{``{q-deformations of two-dimensional Yang-Mills theory: Classification,
  categorification and refinement}''},
\doiref{10.1016/j.nuclphysb.2013.08.001}{Nucl.~Phys. \textbf{B876}, 234
  (2013)\ignorespaces}\ignorespaces,
\normalsize{\texttt{\arxivref{1305.1580}{arXiv:1305.1580}}}\ignorespaces
\bibitem{Buican:2015hsa}
M.~Buican \& T.~Nishinaka,
\textit{``{Argyres-Douglas theories, S$^1$ reductions, and topological
  symmetries}''},
\doiref{10.1088/1751-8113/49/4/045401}{J.~Phys. \textbf{A49}, 045401
  (2016)\ignorespaces}\ignorespaces,
\normalsize{\texttt{\arxivref{1505.06205}{arXiv:1505.06205}}}\ignorespaces
\bibitem{Fredrickson:2017yka}
L.~Fredrickson, D.~Pei, W.~Yan \& K.~Ye,
\textit{``{Argyres-Douglas Theories, Chiral Algebras and Wild Hitchin
  Characters}''},
\normalsize{\texttt{\arxivref{1701.08782}{arXiv:1701.08782}}}\ignorespaces
\bibitem{Cordova:2015nma}
C.~Cordova \& S.-H. Shao,
\textit{``{Schur Indices, BPS Particles, and Argyres-Douglas Theories}''},
\doiref{10.1007/JHEP01(2016)040}{JHEP \textbf{1601}, 040
  (2016)\ignorespaces}\ignorespaces,
\normalsize{\texttt{\arxivref{1506.00265}{arXiv:1506.00265}}}\ignorespaces
\bibitem{Iqbal:2012xm}
A.~Iqbal \& C.~Vafa,
\textit{``{BPS Degeneracies and Superconformal Index in Diverse Dimensions}''},
\doiref{10.1103/PhysRevD.90.105031}{Phys.~Rev. \textbf{D90}, 105031
  (2014)\ignorespaces}\ignorespaces,
\normalsize{\texttt{\arxivref{1210.3605}{arXiv:1210.3605}}}\ignorespaces
\bibitem{Buican:2015tda}
M.~Buican \& T.~Nishinaka,
\textit{``{Argyres-Douglas Theories, the Macdonald Index, and an RG
  Inequality}''},
\doiref{10.1007/JHEP02(2016)159}{JHEP \textbf{1602}, 159
  (2016)\ignorespaces}\ignorespaces,
\normalsize{\texttt{\arxivref{1509.05402}{arXiv:1509.05402}}}\ignorespaces
\bibitem{Song:2015wta}
J.~Song,
\textit{``{Superconformal indices of generalized Argyres-Douglas theories from
  2d TQFT}''},
\doiref{10.1007/JHEP02(2016)045}{JHEP \textbf{1602}, 045
  (2016)\ignorespaces}\ignorespaces,
\normalsize{\texttt{\arxivref{1509.06730}{arXiv:1509.06730}}}\ignorespaces
\bibitem{Cecotti:2015lab}
S.~Cecotti, J.~Song, C.~Vafa \& W.~Yan,
\textit{``{Superconformal Index, BPS Monodromy and Chiral Algebras}''},
\normalsize{\texttt{\arxivref{1511.01516}{arXiv:1511.01516}}}\ignorespaces
\bibitem{Xie:2016evu}
D.~Xie, W.~Yan \& S.-T. Yau,
\textit{``{Chiral algebra of Argyres-Douglas theory from M5 brane}''},
\normalsize{\texttt{\arxivref{1604.02155}{arXiv:1604.02155}}}\ignorespaces
\bibitem{Maruyoshi:2016tqk}
K.~Maruyoshi \& J.~Song,
\textit{``{Enhancement of Supersymmetry via Renormalization Group Flow and the
  Superconformal Index}''},
\doiref{10.1103/PhysRevLett.118.151602}{Phys.~Rev.~Lett. \textbf{118}, 151602
  (2017)\ignorespaces}\ignorespaces,
\normalsize{\texttt{\arxivref{1606.05632}{arXiv:1606.05632}}}\ignorespaces
\bibitem{Maruyoshi:2016aim}
K.~Maruyoshi \& J.~Song,
\textit{``{$ \mathcal{N}=1 $ deformations and RG flows of $ \mathcal{N}=2 $
  SCFTs}''},
\doiref{10.1007/JHEP02(2017)075}{JHEP \textbf{1702}, 075
  (2017)\ignorespaces}\ignorespaces,
\normalsize{\texttt{\arxivref{1607.04281}{arXiv:1607.04281}}}\ignorespaces
\bibitem{Agarwal:2016pjo}
P.~Agarwal, K.~Maruyoshi \& J.~Song,
\textit{``{$ \mathcal{N} $ =1 Deformations and RG flows of $ \mathcal{N} $ =2
  SCFTs, part II: non-principal deformations}''},
\doiref{10.1007/JHEP12(2016)103, 10.1007/JHEP04(2017)113}{JHEP \textbf{1612},
  103 (2016)\ignorespaces}\ignorespaces,
\normalsize{\texttt{\arxivref{1610.05311}{arXiv:1610.05311}}}\ignorespaces,
[Addendum: JHEP04,113(2017)]\ignorespaces
\bibitem{Xie:2016hny}
D.~Xie \& K.~Yonekura,
\textit{``{Search for a Minimal N=1 Superconformal Field Theory in 4D}''},
\doiref{10.1103/PhysRevLett.117.011604}{Phys.~Rev.~Lett. \textbf{117}, 011604
  (2016)\ignorespaces}\ignorespaces,
\normalsize{\texttt{\arxivref{1602.04817}{arXiv:1602.04817}}}\ignorespaces
\bibitem{Buican:2016hnq}
M.~Buican \& T.~Nishinaka,
\textit{``{Small deformation of a simple $\mathcal N=2$ superconformal
  theory}''},
\doiref{10.1103/PhysRevD.94.125002}{Phys.~Rev. \textbf{D94}, 125002
  (2016)\ignorespaces}\ignorespaces,
\normalsize{\texttt{\arxivref{1602.05545}{arXiv:1602.05545}}}\ignorespaces
\bibitem{Cordova:2016uwk}
C.~Cordova, D.~Gaiotto \& S.-H. Shao,
\textit{``{Infrared Computations of Defect Schur Indices}''},
\doiref{10.1007/JHEP11(2016)106}{JHEP \textbf{1611}, 106
  (2016)\ignorespaces}\ignorespaces,
\normalsize{\texttt{\arxivref{1606.08429}{arXiv:1606.08429}}}\ignorespaces
\bibitem{Cordova:2017ohl}
C.~Cordova, D.~Gaiotto \& S.-H. Shao,
\textit{``{Surface Defect Indices and 2d-4d BPS States}''},
\normalsize{\texttt{\arxivref{1703.02525}{arXiv:1703.02525}}}\ignorespaces
\bibitem{Cordova:2017mhb}
C.~Cordova, D.~Gaiotto \& S.-H. Shao,
\textit{``{Surface Defects and Chiral Algebras}''},
\normalsize{\texttt{\arxivref{1704.01955}{arXiv:1704.01955}}}\ignorespaces
\bibitem{Xie:2016uqq}
D.~Xie \& S.-T. Yau,
\textit{``{New N = 2 dualities}''},
\normalsize{\texttt{\arxivref{1602.03529}{arXiv:1602.03529}}}\ignorespaces
\bibitem{Xie:2017vaf}
D.~Xie \& S.-T. Yau,
\textit{``{Argyres-Douglas matter and N=2 dualities}''},
\normalsize{\texttt{\arxivref{1701.01123}{arXiv:1701.01123}}}\ignorespaces
\bibitem{Caorsi:2016ebt}
M.~Caorsi \& S.~Cecotti,
\textit{``{Homological S-Duality in 4d N=2 QFTs}''},
\normalsize{\texttt{\arxivref{1612.08065}{arXiv:1612.08065}}}\ignorespaces
\bibitem{Xie:2013jc}
D.~Xie \& P.~Zhao,
\textit{``{Central charges and RG flow of strongly-coupled N=2 theory}''},
\doiref{10.1007/JHEP03(2013)006}{JHEP \textbf{1303}, 006
  (2013)\ignorespaces}\ignorespaces,
\normalsize{\texttt{\arxivref{1301.0210}{arXiv:1301.0210}}}\ignorespaces
\bibitem{Gaiotto:2009hg}
D.~Gaiotto, G.~W. Moore \& A.~Neitzke,
\textit{``{Wall-crossing, Hitchin Systems, and the WKB Approximation}''},
\normalsize{\texttt{\arxivref{0907.3987}{arXiv:0907.3987}}}\ignorespaces
\bibitem{Gaiotto:2009we}
D.~Gaiotto,
\textit{``{N=2 dualities}''},
\doiref{10.1007/JHEP08(2012)034}{JHEP \textbf{1208}, 034
  (2012)\ignorespaces}\ignorespaces,
\normalsize{\texttt{\arxivref{0904.2715}{arXiv:0904.2715}}}\ignorespaces
\bibitem{Witten:1997sc}
E.~Witten,
\textit{``{Solutions of four-dimensional field theories via M theory}''},
\doiref{10.1016/S0550-3213(97)00416-1}{Nucl.~Phys. \textbf{B500}, 3
  (1997)\ignorespaces}\ignorespaces,
\normalsize{\texttt{\arxivref{hep-th/9703166}{hep-th/9703166}}}\ignorespaces
\bibitem{Gukov:2006jk}
S.~Gukov \& E.~Witten,
\textit{``{Gauge Theory, Ramification, And The Geometric Langlands Program}''},
\normalsize{\texttt{\arxivref{hep-th/0612073}{hep-th/0612073}}}\ignorespaces
\bibitem{Chacaltana:2012zy}
O.~Chacaltana, J.~Distler \& Y.~Tachikawa,
\textit{``{Nilpotent orbits and codimension-two defects of 6d N=(2,0)
  theories}''},
\doiref{10.1142/S0217751X1340006X}{Int.~J.~Mod.~Phys. \textbf{A28}, 1340006
  (2013)\ignorespaces}\ignorespaces,
\normalsize{\texttt{\arxivref{1203.2930}{arXiv:1203.2930}}}\ignorespaces
\bibitem{Witten:2007td}
E.~Witten,
\textit{``{Gauge theory and wild ramification}''},
\normalsize{\texttt{\arxivref{0710.0631}{arXiv:0710.0631}}}\ignorespaces
\bibitem{Bonelli:2011aa}
G.~Bonelli, K.~Maruyoshi \& A.~Tanzini,
\textit{``{Wild Quiver Gauge Theories}''},
\doiref{10.1007/JHEP02(2012)031}{JHEP \textbf{1202}, 031
  (2012)\ignorespaces}\ignorespaces,
\normalsize{\texttt{\arxivref{1112.1691}{arXiv:1112.1691}}}\ignorespaces
\bibitem{Wang:2015mra}
Y.~Wang \& D.~Xie,
\textit{``{Classification of Argyres-Douglas theories from M5 branes}''},
\doiref{10.1103/PhysRevD.94.065012}{Phys.~Rev. \textbf{D94}, 065012
  (2016)\ignorespaces}\ignorespaces,
\normalsize{\texttt{\arxivref{1509.00847}{arXiv:1509.00847}}}\ignorespaces
\bibitem{Dolan:2002zh}
F.~A. Dolan \& H.~Osborn,
\textit{``{On short and semi-short representations for four-dimensional
  superconformal symmetry}''},
\doiref{10.1016/S0003-4916(03)00074-5}{Annals~Phys. \textbf{307}, 41
  (2003)\ignorespaces}\ignorespaces,
\normalsize{\texttt{\arxivref{hep-th/0209056}{hep-th/0209056}}}\ignorespaces
\bibitem{Dobrev:1985qv}
V.~K. Dobrev \& V.~B. Petkova,
\textit{``{All Positive Energy Unitary Irreducible Representations of Extended
  Conformal Supersymmetry}''},
\doiref{10.1016/0370-2693(85)91073-1}{Phys.~Lett. \textbf{B162}, 127
  (1985)\ignorespaces}\ignorespaces
\bibitem{Gadde:2011ia}
A.~Gadde \& W.~Yan,
\textit{``{Reducing the 4d Index to the $S^3$ Partition Function}''},
\doiref{10.1007/JHEP12(2012)003}{JHEP \textbf{1212}, 003
  (2012)\ignorespaces}\ignorespaces,
\normalsize{\texttt{\arxivref{1104.2592}{arXiv:1104.2592}}}\ignorespaces
\bibitem{Nishioka:2011dq}
T.~Nishioka, Y.~Tachikawa \& M.~Yamazaki,
\textit{``{3d Partition Function as Overlap of Wavefunctions}''},
\doiref{10.1007/JHEP08(2011)003}{JHEP \textbf{1108}, 003
  (2011)\ignorespaces}\ignorespaces,
\normalsize{\texttt{\arxivref{1105.4390}{arXiv:1105.4390}}}\ignorespaces
\bibitem{Buican:2016arp}
M.~Buican \& T.~Nishinaka,
\textit{``{Conformal Manifolds in Four Dimensions and Chiral Algebras}''},
\doiref{10.1088/1751-8113/49/46/465401}{J.~Phys. \textbf{A49}, 465401
  (2016)\ignorespaces}\ignorespaces,
\normalsize{\texttt{\arxivref{1603.00887}{arXiv:1603.00887}}}\ignorespaces
\bibitem{Gaiotto:2012xa}
D.~Gaiotto, L.~Rastelli \& S.~S. Razamat,
\textit{``{Bootstrapping the superconformal index with surface defects}''},
\doiref{10.1007/JHEP01(2013)022}{JHEP \textbf{1301}, 022
  (2013)\ignorespaces}\ignorespaces,
\normalsize{\texttt{\arxivref{1207.3577}{arXiv:1207.3577}}}\ignorespaces
\bibitem{DiPietro:2014bca}
L.~Di~Pietro \& Z.~Komargodski,
\textit{``{Cardy formulae for SUSY theories in $d =$ 4 and $d =$ 6}''},
\doiref{10.1007/JHEP12(2014)031}{JHEP \textbf{1412}, 031
  (2014)\ignorespaces}\ignorespaces,
\normalsize{\texttt{\arxivref{1407.6061}{arXiv:1407.6061}}}\ignorespaces
\bibitem{Ardehali:2015bla}
A.~Arabi~Ardehali,
\textit{``{High-temperature asymptotics of supersymmetric partition
  functions}''},
\doiref{10.1007/JHEP07(2016)025}{JHEP \textbf{1607}, 025
  (2016)\ignorespaces}\ignorespaces,
\normalsize{\texttt{\arxivref{1512.03376}{arXiv:1512.03376}}}\ignorespaces
\bibitem{DiPietro:2016ond}
L.~Di~Pietro \& M.~Honda,
\textit{``{Cardy Formula for 4d SUSY Theories and Localization}''},
\normalsize{\texttt{\arxivref{1611.00380}{arXiv:1611.00380}}}\ignorespaces
\bibitem{Buican:2014hfa}
M.~Buican, S.~Giacomelli, T.~Nishinaka \& C.~Papageorgakis,
\textit{``{Argyres-Douglas Theories and S-Duality}''},
\normalsize{\texttt{\arxivref{1411.6026}{arXiv:1411.6026}}}\ignorespaces
\bibitem{DelZotto:2015rca}
M.~Del~Zotto, C.~Vafa \& D.~Xie,
\textit{``{Geometric engineering, mirror symmetry and $
  6{\mathrm{d}}_{\left(1,0\right)}\to
  4{\mathrm{d}}_{\left(\mathcal{N}=2\right)} $}''},
\doiref{10.1007/JHEP11(2015)123}{JHEP \textbf{1511}, 123
  (2015)\ignorespaces}\ignorespaces,
\normalsize{\texttt{\arxivref{1504.08348}{arXiv:1504.08348}}}\ignorespaces
\bibitem{Cecotti:2015hca}
S.~Cecotti \& M.~Del~Zotto,
\textit{``{Higher S-dualities and Shephard-Todd groups}''},
\doiref{10.1007/JHEP09(2015)035}{JHEP \textbf{1509}, 035
  (2015)\ignorespaces}\ignorespaces,
\normalsize{\texttt{\arxivref{1507.01799}{arXiv:1507.01799}}}\ignorespaces
\bibitem{Intriligator:1996ex}
K.~A. Intriligator \& N.~Seiberg,
\textit{``{Mirror symmetry in three-dimensional gauge theories}''},
\doiref{10.1016/0370-2693(96)01088-X}{Phys.~Lett. \textbf{B387}, 513
  (1996)\ignorespaces}\ignorespaces,
\normalsize{\texttt{\arxivref{hep-th/9607207}{hep-th/9607207}}}\ignorespaces
\bibitem{DelZotto:2014kka}
M.~Del~Zotto \& A.~Hanany,
\textit{``{Complete Graphs, Hilbert Series, and the Higgs branch of the 4d
  $\mathcal{N} =$ 2 $(A_n,A_m)$ SCFTs}''},
\doiref{10.1016/j.nuclphysb.2015.03.017}{Nucl.Phys. \textbf{B894}, 439
  (2015)\ignorespaces}\ignorespaces,
\normalsize{\texttt{\arxivref{1403.6523}{arXiv:1403.6523}}}\ignorespaces
\bibitem{Gaiotto:2008ak}
D.~Gaiotto \& E.~Witten,
\textit{``{S-Duality of Boundary Conditions In N=4 Super Yang-Mills Theory}''},
\doiref{10.4310/ATMP.2009.v13.n3.a5}{Adv.~Theor.~Math.~Phys. \textbf{13}, 721
  (2009)\ignorespaces}\ignorespaces,
\normalsize{\texttt{\arxivref{0807.3720}{arXiv:0807.3720}}}\ignorespaces
\bibitem{deBoer:1996ck}
J.~de~Boer, K.~Hori, H.~Ooguri, Y.~Oz \& Z.~Yin,
\textit{``{Mirror symmetry in three-dimensional theories, SL(2,Z) and D-brane
  moduli spaces}''},
\doiref{10.1016/S0550-3213(97)00115-6}{Nucl.~Phys. \textbf{B493}, 148
  (1997)\ignorespaces}\ignorespaces,
\normalsize{\texttt{\arxivref{hep-th/9612131}{hep-th/9612131}}}\ignorespaces
\bibitem{Gaiotto:2014kfa}
D.~Gaiotto, A.~Kapustin, N.~Seiberg \& B.~Willett,
\textit{``{Generalized Global Symmetries}''},
\doiref{10.1007/JHEP02(2015)172}{JHEP \textbf{1502}, 172
  (2015)\ignorespaces}\ignorespaces,
\normalsize{\texttt{\arxivref{1412.5148}{arXiv:1412.5148}}}\ignorespaces
\bibitem{Seiberg:1994aj}
N.~Seiberg \& E.~Witten,
\textit{``{Monopoles, duality and chiral symmetry breaking in N=2
  supersymmetric QCD}''},
\doiref{10.1016/0550-3213(94)90214-3}{Nucl.~Phys. \textbf{B431}, 484
  (1994)\ignorespaces}\ignorespaces,
\normalsize{\texttt{\arxivref{hep-th/9408099}{hep-th/9408099}}}\ignorespaces
\bibitem{Mekareeya:2012tn}
N.~Mekareeya, J.~Song \& Y.~Tachikawa,
\textit{``{2d TQFT structure of the superconformal indices with
  outer-automorphism twists}''},
\doiref{10.1007/JHEP03(2013)171}{JHEP \textbf{1303}, 171
  (2013)\ignorespaces}\ignorespaces,
\normalsize{\texttt{\arxivref{1212.0545}{arXiv:1212.0545}}}\ignorespaces
\bibitem{Lemos:2012ph}
M.~Lemos, W.~Peelaers \& L.~Rastelli,
\textit{``{The superconformal index of class $S$ theories of type $D$}''},
\doiref{10.1007/JHEP05(2014)120}{JHEP \textbf{1405}, 120
  (2014)\ignorespaces}\ignorespaces,
\normalsize{\texttt{\arxivref{1212.1271}{arXiv:1212.1271}}}\ignorespaces
\bibitem{Cordova:2015vwa}
C.~Cordova, T.~T. Dumitrescu \& X.~Yin,
\textit{``{Higher Derivative Terms, Toroidal Compactification, and Weyl
  Anomalies in Six-Dimensional (2,0) Theories}''},
\normalsize{\texttt{\arxivref{1505.03850}{arXiv:1505.03850}}}\ignorespaces
\bibitem{Shimizu:2017kzs}
H.~Shimizu, Y.~Tachikawa \& G.~Zafrir,
\textit{``{Anomaly matching on the Higgs branch}''},
\normalsize{\texttt{\arxivref{1703.01013}{arXiv:1703.01013}}}\ignorespaces
\bibitem{Creutzig:2017qyf}
T.~Creutzig,
\textit{``{W-algebras for Argyres-Douglas theories}''},
\normalsize{\texttt{\arxivref{1701.05926}{arXiv:1701.05926}}}\ignorespaces
\bibitem{BR}
C.~Beem \& L.~Rastelli,
\textit{``{To Appear}''}
\bibitem{Argyres:2016xua}
P.~C. Argyres, M.~Lotito, Y.~Lu \& M.~Martone,
\textit{``{Expanding the landscape of $\mathcal N = 2$ rank 1 SCFTs}''},
\doiref{10.1007/JHEP05(2016)088}{JHEP \textbf{1605}, 088
  (2016)\ignorespaces}\ignorespaces,
\normalsize{\texttt{\arxivref{1602.02764}{arXiv:1602.02764}}}\ignorespaces
\end{thebibliography}

\end{document}